\documentclass[%
 twocolumn,
 superscriptaddress,
 preprintnumbers,
 nofootinbib,
 amsmath,amssymb,
 aps,
 prl,
]{revtex4}
\usepackage{psfrag,slashed,cancel,array,graphicx,hyperref}
\usepackage{subfigure}
\usepackage[utf8]{inputenc}
\usepackage{mathtools}
\usepackage{mathrsfs}
\usepackage{multirow}
\usepackage{dblfloatfix}
\usepackage{braket}
\usepackage{booktabs} 
\usepackage[normalem]{ulem}
\hypersetup{pdftitle={},pdfcreator={},linkcolor=[rgb]{0.15,0.35,0.75},colorlinks=true,citecolor=[rgb]{0.675,0,0.2},urlcolor=[rgb]{0.15,0.35,0.65}}

\newcommand{\um}{\mathfrak{u}}

\newcommand{\bm}{\mathfrak{b}}

\allowdisplaybreaks

\begin{document}


\title{NNLO QCD corrections to unpolarized and polarized electroweak structure functions in semi-inclusive deep-inelastic scattering}

\author{Saurav Goyal}
\email{sauravg@imsc.res.in}
\affiliation{The Institute of Mathematical Sciences,  Taramani, 600113 Chennai, India}
\affiliation{Homi Bhabha National Institute, Training School Complex, Anushakti Nagar, Mumbai 400094, India}
\author{Sven-Olaf Moch}
\email{sven-olaf.moch@desy.de}
\affiliation{II. Institute for Theoretical Physics, Hamburg University, D-22761 Hamburg, Germany} 
\author{Vaibhav Pathak}
\email{vaibhavp@imsc.res.in}
\affiliation{The Institute of Mathematical Sciences, Taramani, 600113 Chennai, India}
\affiliation{Homi Bhabha National Institute, Training School Complex, Anushakti Nagar, Mumbai 400094, India}
\author{V. Ravindran}
\email{ravindra@imsc.res.in}
\affiliation{The Institute of Mathematical Sciences, Taramani, 600113 Chennai, India}
\affiliation{Homi Bhabha National Institute, Training School Complex, Anushakti Nagar, Mumbai 400094, India}

\date{\today}

\begin{abstract}
We present results for unpolarized and polarized semi-inclusive deep-inelastic scattering mediated by electroweak gauge bosons at next-to-next-to-leading order (NNLO) in perturbative quantum chromodynamics. The results include all relevant structure functions arising from both neutral current (NC) and charged current (CC) interactions, incorporating  contributions from all partonic channels with full flavor dependence. These corrections are crucial for improving the theoretical precision. A detailed numerical analysis of the NNLO corrections demonstrates their phenomenological importance, revealing sizable effects and a significant reduction in residual scale dependence in the kinematic range probed by the future Electron-Ion-Collider. These results will serve as a critical input for future global extractions of parton distributions functions and fragmentation functions.
\end{abstract}
 


\maketitle


One of the central challenges in high-energy particle physics is to achieve a comprehensive understanding of the internal structure of the proton and other hadrons in terms of their constituents such as quark and gluon degrees of freedom. In particular, 
accurately extracting the parton distribution functions (PDFs) to understand the momentum distributions of quarks and gluons remains a fundamental objective. Equally important is understanding the spin structure of the proton, namely how the total nucleon spin arises from the intrinsic spin and orbital angular momentum of quarks and gluons.
Disentangling these contributions requires detailed knowledge of the flavor, polarization, and momentum dependence of partons across a wide range of kinematic regimes. 

In high energy colliders, the production of hadrons in the final state can be studied with the help of  fragmentation functions (FFs).  The FFs parameterize the transition of energetic partons (quarks or gluons) into observable hadrons through the nonperturbative dynamics of quantum chromodynamics (QCD). They describe the probability that a parton produced in a short-distance scattering process fragments into a particular hadron carrying a fraction $z$ of the parent parton’s momentum.  
A precise determination of these functions is thus essential, as it offers key insights into the hadronization mechanism, the process by which colored partons from high-energy reactions transform into color-neutral hadrons. 
Consequently, improving our knowledge of FFs is crucial for both precision QCD phenomenology and a deeper understanding of the nonperturbative dynamics responsible for hadron formation. 
Studying flavor-changing and flavor-preserving interactions in SIDIS enhances the sensitivity of  both  (unpolarized/polarized) PDFs and FFs~\cite{deFlorian:2017lwf,Anderle:2016czy,Leader:2015hna,Bertone:2018ecm,Khalek:2021gxf,Ethier:2017zbq,DeFlorian:2019xxt,Moffat:2021dji,Alekhin:2017kpj,Hou:2019efy,Bailey:2020ooq,NNPDF:2021njg,Alekhin:2024bhs,Metz:2016swz}. 
Identifying final-state hadrons provides additional flavor information and insight into how quarks form hadrons, enabling precise determinations of polarized and unpolarized structure functions (SFs) and disentangling contributions from different quark flavors.

 At the same time, precision studies require an equally precise understanding of the perturbatively calculable short-distance contributions to hadronic scattering. Within the QCD factorization framework, observable cross sections are expressed as convolutions of universal nonperturbative functions, such as PDFs and FFs, with short-distance coefficient functions (CFs) that describe the underlying partonic scattering reactions. These CFs arise from hard parton-level interactions and can be systematically computed order by order in the strong coupling using perturbative QCD. Higher-order corrections in the strong coupling coming from NC and CC mediated processes are now essential to achieve the level of theoretical accuracy required by modern experiments.

Deep-inelastic scattering (DIS)~\cite{Moch:2004xu,Vermaseren:2005qc,Moch:2008fj,Blumlein:2022gpp}, has played a fundamental role in revealing the partonic structure of the nucleon. In DIS processes, high-energy leptons scatter off hadrons through the exchange of a highly virtual photon or weak gauge boson, thereby probing the internal quark and gluon degrees of freedom at short distances. Measurements of inclusive DIS SFs over a wide range of the Bjorken scaling variable $x$ and momentum transfer $Q^2$ 
have provided crucial information on the PDFs. 
Experiments at fixed-target labs and the HERA collider have mapped PDFs across broad ranges of 
$(x,Q^2)$, enabling precise QCD tests via scaling violations and PDF evolution. These measurements remain central to global PDF analyses for high-energy hadronic predictions.

Semi-inclusive deep-inelastic scattering (SIDIS) provides a powerful framework to probe this rich internal structure, as it allows simultaneous access to PDFs and FFs. Achieving reliable phenomenological interpretations of SIDIS measurements therefore requires highly precise theoretical predictions, including higher-order QCD corrections, electroweak effects, and their interplay. Such developments are essential for improving global analyses of PDFs and FFs and for advancing our understanding of the momentum and spin structure of the proton and nuclei. 

The upcoming Electron-Ion Collider (EIC)  will explore the internal structure of nucleons and nuclei with unprecedented precision by colliding polarized electrons with polarized protons and ions over a wide kinematic range in ($x, Q^2$). Its goals include mapping the three-dimensional spatial, momentum, and spin structure of hadrons and significantly improving knowledge of both unpolarized and polarized PDFs. By detecting identified hadrons in the final state, SIDIS provides simultaneous sensitivity to PDFs and FFs and is therefore one of the most valuable observables at the EIC \cite{Aschenauer:2019kzf,Accardi:2012qut,AbdulKhalek:2021gbh}.
At sufficiently high $Q^2$, the EIC will also probe electroweak effects beyond virtual photon exchange. Contributions from $Z$-boson exchange in NC interactions and $W^\pm$-boson exchange in CC interactions will become relevant, adding new SFs and couplings to the cross section. These effects open new opportunities for studying flavor decomposition, quark helicity distributions, and parity-violating asymmetries, and will provide additional constraints on PDFs and FFs in a flavor-sensitive way~\cite{deFlorian:2017lwf,Anderle:2016czy,Leader:2015hna,Bertone:2018ecm,Khalek:2021gxf,Ethier:2017zbq,DeFlorian:2019xxt,Moffat:2021dji,Alekhin:2017kpj,Hou:2019efy,Bailey:2020ooq,NNPDF:2021njg,Alekhin:2024bhs,Metz:2016swz,Bertone:2024taw,Borsa:2024mss,deFlorian:2009vb,deFlorian:2014yva}.

\if{1=0}
The upcoming Electron-Ion Collider (EIC) is 
a next-generation facility 
designed to explore the internal structure of nucleons and nuclei and their dynamics with unprecedented precision. 
By colliding polarized electrons with polarized protons and ions over a wide range of center-of-mass energies, the EIC will provide new insights into how the spin of hadrons arises from their quark and gluon constituents. One of its primary scientific goals is to map the three-dimensional structure of hadrons, including the spatial, momentum, and spin distributions of their partonic constituents. Through measurements covering a broad kinematic range in $(x,Q^2)$.  It is expected that the EIC will significantly improve our knowledge of both unpolarized and polarized {PDFs}.  In addition by detecting identified hadrons in the final state, 
the SIDIS provides simultaneous sensitivity to  PDFs  and  FFs, allowing a more detailed investigation of parton dynamics and hadronization mechanisms.  This makes SIDIS~\cite{Aschenauer:2019kzf} the most promising and valuable observable to probe PDFs as well as FFs at EIC~\cite{Accardi:2012qut,AbdulKhalek:2021gbh}. 

Operating at sufficiently high momentum transfer $Q^2$, the EIC will probe electroweak effects beyond virtual photon exchange. In particular, $Z$-boson exchange through NC interactions and $W^\pm$-boson exchange through CC interactions will contribute, opening the possibility for studies with unprecedented theoretical precision. 
The presence of these interactions not only contributes to SFs and couplings in the cross section, but also open new avenues for studying flavor decomposition, quark helicity distributions, and parity-violating asymmetries. 
These measurements will also help to constrain both PDFs and FFs in a flavor-sensitive manner~\cite{deFlorian:2017lwf,Anderle:2016czy,Leader:2015hna,Bertone:2018ecm,Khalek:2021gxf,Ethier:2017zbq,DeFlorian:2019xxt,Moffat:2021dji,Alekhin:2017kpj,Hou:2019efy,Bailey:2020ooq,NNPDF:2021njg,Alekhin:2024bhs,Metz:2016swz,Bertone:2024taw,Borsa:2024mss,deFlorian:2009vb,deFlorian:2014yva}.
\fi

Recently, there have been several studies aimed at improving theoretical predictions for SIDIS within perturbative QCD.  The earliest work that deals with the next-to-leading order (NLO) QCD corrections to CFs for SIDIS appeared long ago~\cite{Altarelli:1979kv}, see also
\cite{Baier:1979sp,Nason:1993xx,Furmanski:1981cw,Graudenz:1994dq,deFlorian:2012wk}.
Results for NC CFs of $F_{1,2}$ and $g_{1}$ up to next-to-next-to-leading order (NNLO) accuracy in QCD with photon exchange have become available~\cite{Goyal:2023zdi,Bonino:2024qbh,Bonino:2024wgg,Goyal:2024tmo,Goyal:2024emo,Bonino:2024adk}.
Very recently, pure QED and mixed QCD$\otimes$QED corrections up to NNLO level have become available for photon exchange SIDIS, see~\cite{Goyal:2025qyu}.  A comprehensive study of both unpolarized and polarized SFs in QCD up to NNLO level for CC and NC reactions has been accomplished in ~\cite{Bonino:2025tnf,Bonino:2025qta,Bonino:2025bqa}.

In the present work, we revisit these processes within an independent computational framework and analysis setup. This provides an important cross-check of existing results and offers complementary insights into the perturbative structure of SIDIS observables.

We consider the SIDIS process $l(k_l,\lambda_{l}) + H(P,\lambda_{H}) \rightarrow  l'({k}_{l'}) + H'(P_H) + X$  where $l$ ($l'$) and $H$ ($H'$) denote the incoming (outgoing) lepton and hadron respectively.   For NC {processes}, we have $l'=l$ and the intermediate virtual {particles} are $\gamma^*$, $Z^*$ and interference between them. For the CC processes , the intermediate vector boson is either $W^+$ or $W^-$ depending on the charge of $l'$ ($l'\not =l$).
Here $k_l$ ( $k_{l'}$) and $P$ ($P_H$) are momenta of incoming (outgoing) lepton and hadron respectively and $q= k_l - k_{l'}$ is the momentum transfer between incoming leptons and hadrons.  We define, $q^2=-Q^2$. $\lambda_{l}$ and $\lambda_{H}$ are the helicity of incoming lepton  and hadron respectively.
More on these processes can be found in ~\cite{Yang:2020qsk,Chen:2020ugq,Moreno:2014kia,deFlorian:2012wk,Anselmino:2001ey,Boer:1999uu,Anselmino:1994gn,Aschenauer:2013iia}.

In the single-vector-boson exchange approximation, the differential cross section factorizes into a leptonic tensor $L^{\mu\nu}_{j,\lambda_l}$ and a hadronic tensor $W^{j}_{\mu\nu,\lambda_H}$:
\begin{align}
\label{eq:diffcross}
\frac{d^3 \sigma_{\lambda_l\lambda_{H}}}{dxdydz} &= 2\pi y M_{H} \alpha_e^{2} \nonumber\\
\times& \sum_{j} D_{j} L^{\mu\nu}_{j,\lambda_{l}}(k_l,k_{l'},q) W_{\mu\nu,\lambda_{H}}^{j}(P,P_H,S,q)\, .
\end{align}
Here, $M_H$ and $S$ are the mass and spin
of the incoming hadron, $x=\frac{Q^2}{2 P \cdot q}$ is the  Bjorken variable, 
$y=\frac{P\cdot q}{P\cdot k_l}$, the inelasticity and 
$z= \frac{P \cdot P_H}{P \cdot q}$ is  the fraction of the initial energy
transferred to the final-state hadron. $\alpha_e$ is the fine structure constant, $\alpha_e = g_e^2/(4 \pi)$ and $g_e$ is the electromagnetic coupling. 
For NC processes the sum includes contributions from   $j =\gamma\gamma, ZZ$, and $j=\gamma Z$ intermediate vector bosons, while for CC processes $j = WW$. 
Here,
\begin{align}
&D_{\gamma\gamma} = \frac{1}{(Q^2)^2}\, ,\qquad
D_{\gamma Z} = \frac{(Q^2 + M_Z^2)}{Q^2((Q^2 + M_Z^2)^2+\Gamma_Z^2)}\, , \nonumber\\
&D_{Z Z} = \frac{1}{((Q^2 + M_Z^2)^2+\Gamma_Z^2)}\, ,\nonumber\\
&D_{W W } = \frac{1}{((Q^2 + M_W^2)^2+\Gamma_W^2)}
\, ,
\end{align}
where $M_V$ and $\Gamma_V$ ($V=Z,W$) are the mass and width of the intermediate vector boson $V$.

For NC interactions, $\gamma$ and $Z$ couple to a fermion of flavor $f$ through the vertex
\begin{align}
-i g_e e_f\,\gamma^\mu,~~i g_e\,\gamma^\mu\left(\overline{g}_{V_f} -\overline{g}_{A_f}\gamma^5\right),
\end{align}
receptively, where
\begin{align}\label{eq:gAgVbar}
\overline{g}_{V_f} =  \frac{g_{V_f}}{2 \sin\theta_W \cos\theta_W},~~\overline{g}_{A_f} &=  \frac{g_{A_f}}{2 \sin\theta_W \cos\theta_W}\,.
\end{align}
Here $\theta_W$ is the Weinberg mixing angle, and the vector ($g_{V_f}$) and axial-vector ($g_{A_f}$) couplings are given by
\begin{align}\label{eq:gAgV}
g_{V_f}&=T_f^3-2 e_f \sin^2\theta_W, \qquad g_{A_f}=T_f^3.
\end{align}
$T_f^3$ denotes the third component of the weak isospin of the fermion $f$, 
and $e_f$ is the electric charge of the fermion $f$ in units of the electron charge~\cite{ParticleDataGroup:2024cfk}.

For CC interactions, the vector bosons $W^+$ and $W^-$ couple to fermions through the vertices
\begin{align}\label{eq:Vffbar}
i g_e\gamma^\mu\left({1-\gamma^5}\right)\overline{V}_{ff'},
~~
i g_e\gamma^\mu\left({1-\gamma^5}\right)\overline{V}_{ff'}^{*}.
\end{align}
respectively, $\overline{V}_{ff'} = \frac{V_{ff'}}{2\sqrt{2}\sin\theta_W}$ 
where $f$ and $f'$ denote the members of a weak isospin doublet. 
In the quark sector, $f=\um$ represents an up-type quark ($u,c$) and $f'=\bm$ represents a down-type quark ($d,s$), with $V_{\um\bm}$ being the corresponding element of the Cabibbo-Kobayashi-Maskawa (CKM) matrix. In the leptonic sector, $f=l$ and $f'=\nu_l$ correspond to a charged lepton and its associated neutrino ($l=e,\mu,\tau$), respectively. In this case the coupling is governed by the Pontecorvo-Maki-Nakagawa-Sakata (PMNS) matrix. However, since the flavor of the incoming lepton is fixed, the PMNS mixing typically does not appear explicitly in the cross section.

 To define the helicity of fermions we used the relation 
\begin{align}
    u(p,{\lambda})  \overline{u}(p,{\lambda}) = \left(\frac{1+\lambda \gamma^{5}}{2}\right)\slash \!\!\!p
\end{align}
for fermions with $\lambda =1$  for $+$ or $\uparrow$ helicity and $\lambda =-1$  for $-$  or $\downarrow$ helicity.

The leptonic tensors $L^{\mu\nu}_{j,\lambda_l}$ corresponding to exchanges of different
electroweak bosons, $j=\gamma \gamma,ZZ,WW$, are found to be
\begin{align}
\label{eq:lepten}
L^{\mu\nu}_{\gamma \gamma,\lambda_{l}} &=
2k_{l}^{\mu}k_{l'}^{\nu}
+2k_{l'}^{\mu}k_{l}^{\nu}
-Q^2 g^{\mu\nu}
-2il_l\lambda_{l}\epsilon^{\mu\nu\sigma\lambda}q_{\sigma}k_{l,\lambda}, \nonumber\\[4pt]
L^{\mu\nu}_{Z Z,\lambda_{l}} &=
(\overline{g}_{V_l} - l_l\lambda_{l} \overline{g}_{A_l})^2
\,L^{\mu\nu}_{\gamma\gamma,\lambda_{l}}, \nonumber\\[4pt]
L^{\mu\nu}_{\gamma Z,\lambda_{l}} &=
 (\overline{g}_{V_l} - l_l\lambda_{l} \overline{g}_{A_l})
\,L^{\mu\nu}_{\gamma\gamma,\lambda_{l}}, \nonumber\\[4pt]
L^{\mu\nu}_{W W,\lambda_{l}} &=
(1 - l_l\lambda_l)^2
\,L^{\mu\nu}_{\gamma\gamma,\lambda_{l}} .
\end{align}
Here $\lambda_l,l_l=\pm1$ represents the helicity, lepton number of the incoming lepton respectively.
The  Levi-Civita tensor $\epsilon^{\mu\nu\sigma\lambda}$ is fully anti-symmetric tensor with $\epsilon_{0123} = -\epsilon^{0123} = 1$.

The hadronic tensor $W_{\mu\nu,\lambda_H}^{j}$ can be decomposed in terms of
Lorentz tensors $T_{i,\mu\nu}$ and the corresponding SFs.
For SIDIS it takes the form
\begin{widetext}
\begin{align}
\label{eq:hadten}
W_{\mu\nu,\lambda_H}^{j} =& \frac{1}{M_{H}}\Bigg\{ 
F_1^{j}~T_{1,\mu\nu}+ F_2^{j}~T_{2,\mu\nu} - i \epsilon_{\mu\nu\sigma\lambda} \frac{q^{\sigma} P^\lambda}{2P\cdot q} F_3^{j} 
+ 
i \epsilon_{\mu\nu\sigma\lambda} \frac{q^\sigma }{P\cdot q} \left[ S^\lambda g_{1}^{j} + \left({S}^\lambda -\frac{S\cdot q}{P\cdot q} {P}^\lambda \right) g_{2}^{j}\right] 
\nonumber \\&
+ \frac{1}{P\cdot q} g_{3}^{j}\left[ \frac{1}{2} \left(\hat{P}_\mu \hat{S}_\nu + \hat{S}_\mu \hat{P}_\nu\right) - \frac{S\cdot q}{P\cdot q} \hat{P}_\mu \hat{P}_\nu \right]  
+ \frac{S\cdot q}{P\cdot q} \left[  g_{4}^{j}~T_{2,\mu\nu} + g_{5}^{j}~T_{1,\mu\nu} \right] \Bigg\} \,,
\end{align}    
with
\begin{align}
\hat{P}_\mu=& P_\mu - \frac{P\cdot q}{q^2}q_\mu \, , &
\hat{S}_\mu=& S_\mu - \frac{S\cdot q}{q^2}q_\mu\,,
\end{align}
%
where $F_{1,2,3}^{j}=F_{1,2,3}^{j}(x,z,Q^2)$ denote the unpolarized
SFs, while $g_{1,4,5}^{j}=g_{1,4,5}^{j}(x,z,Q^2)$ correspond
to the polarized ones and $S$ is the spin of incoming hadron aligned in longitudinal direction of lepton $l$,  such that in $S$ = $\lambda_{H}(0,0,0,M_{H})$, in rest frame of incoming hadron and  $S\!\cdot\!P=0$ and $S^2=-M_{H}^2$.
The Lorentz tensors $T_{1,2}^{\mu\nu}$ are defined as
\begin{align}
T_{1,\mu\nu} = -g_{\mu\nu} + \frac{q_\mu q_\nu}{q^2}\, , \qquad
T_{2,\mu\nu} = \frac{1}{P\!\cdot\!q}
\left(P_\mu - \frac{P\!\cdot\!q}{q^2}q_\mu\right)
\left(P_\nu - \frac{P\!\cdot\!q}{q^2}q_\nu\right).
\end{align}
Substituting Eqs.~\eqref{eq:lepten} and \eqref{eq:hadten} in Eq.~\eqref{eq:diffcross}, we find differential cross section,
%
\begin{align}
\label{eq:upol}
\frac{d^3 \sigma^{\text{I}}_{\lambda_{l}\lambda_{H}}}{dxdydz}
&=
\frac{4\pi\alpha_e^2}{Q^2}
\Bigg[
y~F^{\text{I}}_1 + \bigg(\frac{1-y}{xy} -  x y\frac{M_{H}^2}{Q^2}\bigg) F^{\text{I}}_2 +\big(2-y\big)\frac{F^{\text{I}}_3}{2}
\nonumber\\
&\quad
+ \lambda_{H}\Bigg\{\bigg(2-y -2x^2y^2\frac{M_{H}^2 }{Q^2}\bigg)g_{1}^{\text{I}}
+ 4x^2y\frac{M_{H}^2}{Q^2}~g_{2}^{\text{I}} + 2x\frac{M_{H}^2}{Q^2}\bigg(1-y-x^2y^2\frac{M_{H}^2}{Q^2}\bigg)g_{3}^{\text{I}}
\nonumber\\
&\quad+
\bigg(1 + 2 x^2 y\frac{M_{H}^2}{Q^2}\bigg)\bigg[ \bigg(\frac{1-y}{xy} - x y\frac{M_{H}^2 }{Q^2}\bigg)g_{4}^{\text{I}} 
+ y~g_{5}^{\text{I}}
\bigg]
\Bigg\}
\Bigg]\,,
%
\end{align}    
%
where I = NC, CC. 
The NC SFs can be decomposed into the $\gamma\gamma$, $ZZ$, and $\gamma Z$  contributions as,
%
\begin{align}
\label{eq:pol}
F^{\text{NC}}_{1,2}
&= \eta_{\gamma\gamma} F^{\gamma\gamma}_{1,2}
  + \eta_{ZZ}
  \left(g_{V_l}^2 + g_{A_l}^2- 2l_{l}\lambda_{l} g_{V_l} g_{A_l} \right) 
  F^{ZZ}_{1,2} + \eta_{\gamma Z}\left(g_{V_l}-l_{l}\lambda_{l}g_{A_{l}}\right) F^{\gamma Z}_{1,2},
\\
F^{\text{NC}}_{3}
&= -\eta_{\gamma \gamma}(l_{l}\lambda_{l}) F^{\gamma \gamma}_{3} -
\eta_{ZZ} 
~l_{l}\lambda_{l}
  \left(g_{V_l}^2 + g_{A_l}^2- 2l_{l}\lambda_{l} g_{V_l} g_{A_l} \right) 
F^{ZZ}_{3}
- \eta_{\gamma Z}~l_{l}\lambda_{l}\left(g_{V_l}-l_{l}\lambda_{l}g_{A_l}\right) F^{\gamma Z}_{3}.
\end{align}
\end{widetext}
For CC interactions, the intermediate vector boson is either $W^+$ or $W^-$ leading to
the simple relation
\begin{align}
\label{eq:pol}
F^{\text{CC}}_{1,2,3}
&= \eta_{WW}~(1 - l_{l}\lambda_{l}  )^2F^{WW}_{1,2,3}\,.
\end{align}
 From now on we'll consider $Q^2 \gg M_{H}^2$, such that all the hadron mass dependent terms in Eq.~(\ref{eq:upol}) and also the SFs $g_{2}^{\text{I}}$ and $g^{\text{I}}_{3}$ can be dropped.
For the polarized case, the decomposition is identical to that of unpolarized ones i.e.,
%
$g^{\text{NC/CC}}_{1}
= -l_{l}\lambda_{l}\times F^{\text{NC/CC}}_{1}\Big|_{F^j_1 \rightarrow g^j_1}\,\, \text{and} \,\,
g^{\text{NC/CC}}_{4,5}= l_{l}\lambda_{l}\times F^{\text{NC/CC}}_{3}\Big|_{F^j_{3} \rightarrow g^j_{4,5}}$.
The constants $\eta_j$ encode the electroweak factors associated with the exchanged bosons. They are given by
\begin{align}
\eta_{\gamma \gamma } &= 1 \,, \\
\eta_{\gamma  Z } &= 
\sqrt{\frac{\sqrt{2} G_F M_Z^2}{4\pi \alpha_e}}
\frac{Q^2(Q^2+M_Z^2)}{(Q^2+M_Z^2)^2+\Gamma_Z^2} \,, \\
\eta_{Z Z} &= \left(\frac{\sqrt{2}G_F M_Z^2}{4\pi \alpha_e}\right)
\frac{(Q^2)^2}{(Q^2+M_Z^2)^2+\Gamma_Z^2} \,, \\
\eta_{WW} &= 
\left(
\frac{G_F M_W^2}{\sqrt{2}(4\pi \alpha_e)}\right)
\frac{(Q^2)^2}{(Q^2+M_W^2)^2+\Gamma_W^2}
 .
\end{align}
Here $G_F$ denotes the Fermi constant.

Since the tensors $T_{1,\mu\nu}$ and $T_{2,\mu\nu}$ are symmetric in the
indices $\mu$ and $\nu$, the corresponding unpolarized SFs
$F^{\text{NC}}_{1,2}$ receive contributions only from the vector-vector
and axial-axial parts of the NC. If the incoming lepton is polarized, then the SFs $g^{\text{NC}}_{4,5}$ will receive contribution from interference of vector-axial vector parts. In contrast, 
due to  anti-symmetric nature of Levi-Civita tensor,
the SF $F^{\text{NC}}_{3}$ arises from the interference of vector and axial-vector components of the NC for the unpolarized incoming lepton.  When the incoming lepton is polarized, then the vector-vector and axial vector-axial vector parts will also contributes in $g^{\text{NC}}_{1}$.

The SFs can be expressed in terms of perturbatively calculable CFs $(\Delta)\mathcal{C}_{i,ab}^j$, the PDF $(\Delta)f_{a/H}$ of a parton $a$ inside the incoming hadron $H$, and the FF $D_{H'/b}$ describing the fragmentation of parton $b$ into the outgoing hadron $H'$, 
where the symbol $(\Delta)$ unifies the notation for unpolarized and polarized quantities. 
\begin{widetext}
\begin{align}
(\Delta)\mathcal{F}_i^{\,j}(x,z,Q^2)
&= \sum_{a,b}
\int_x^1 \frac{dx_1}{x_1}\,
(\Delta)f_{a/H}(x_1,\mu_F^2)
\int_z^1 \frac{dz_1}{z_1}\,
D_{H'/b}(z_1,\mu_F^2) 
(\Delta)\mathcal{C}_{i,ab}^{\,j}
\left(
\frac{x}{x_1},\frac{z}{z_1},
\mu_F^2,Q^2
\right).
\label{eq:SFdef}
\end{align}
\end{widetext}
Unpolarized SFs take indices $i = 1,2,3$ and polarized ones $i = 1,4,5$,
%
\begin{align}
\mathcal{F}_{1,2,3}^j=\left\{2F_1^j,\frac{F_2^j}{x},F_3^j\right\}\, , 
\quad
\Delta\mathcal{F}_{1,4,5}^j=\left\{g_1^j,\frac{g_4^j}{2x},g_5^j \right\}\, .
\end{align}
In Eq.~\eqref{eq:SFdef} the sum is over all possible incoming and outgoing partons $a,b=q,\overline{q}',g$ and $\mu_F$ is the factorization scale. The scaling variable  $x_1$ is the momentum fraction carried away by the parton $a$ 
of the incident hadron $H$ such that ${p_a} = x_1{P}$,  and $z_1$ is the momentum fraction carried away by the final hadron $H'$ from outgoing parton $b$, i.e. ${P_{H}}=z_{1}{p_b}$.
Here, $p_a$ and $p_b$ are momenta of incident parton and fragmenting parton respectively.  
The  CFs ${(\Delta) \cal C}^{j}_{i,ab}$ are computable in perturbative QCD  
in powers of $a_s(\mu_R^2)$,
\begin{align}    
\label{eq:as-exp}
(\Delta){\cal C}^{j}_{i,ab}(\mu_F^2) = \sum_{k=0}^\infty\, a_s^k(\mu_R^2)\, (\Delta) {\cal C}_{i,ab}^{j,(k)}(\mu_F^2,\mu_R^2)
\, ,
\end{align}
where
$a_s(\mu_R^2)= {g_s^2(\mu_R^2)}/(16\pi^2)$ in terms of the strong coupling $g_s$, defined in $\rm{\overline {MS}}$ scheme at the renormalization scale $\mu_R$. 
The CFs $(\Delta)\mathcal{C}^{\,j}_{i,ab}$ are obtained from
the partonic subprocess cross sections 
$d^3(\Delta)\hat{\sigma}^{\,j}_{i,ab}/(dx\,dy\,dz)$ 
(or $d(\Delta)\hat{\sigma}^{\,j}_{i,ab}$ for short) by applying the
corresponding projectors $(\Delta)\mathcal{P}^{\mu\nu}_{i}$. 
\begin{align}
\label{eq:parton-crs}
d(\Delta) \hat \sigma^j_{i,ab} = & \frac{(\Delta)\mathcal{P}_{i}^{\mu\nu}}{4\pi}  \int  \text{dPS}_{X+b}\,  (\Delta) \overline{\Sigma}|{M}_{ab}|^{2}_{j,\mu\nu}\, 
\nonumber \\
& \times \delta\Big(\frac{z}{z_1}-\frac{p_a \cdot p_b}{p_a\cdot q}\Big)\, ,
\end{align}
where $(\Delta) M_{ab} = M_{a(\uparrow)b}+(-)M_{a(\downarrow)b}$ is the spin-independent (dependent) amplitude for the process,
\begin{align}
a(p_a,s_a) + V^*(q) \;\to\; b(p_b) + X,
\end{align}
where the parton $b$ fragments into the hadron $H'$. 
The spin label $s_a$ appears only for polarized subprocesses, while in the 
unpolarized case the spin is averaged over and the dependence on $s_a$ is suppressed.
The subscript $j$ labels the type of intermediate bosons: 
$j = V {V'}$, corresponding to $\gamma\gamma$, $ZZ$, or $\gamma Z$, 
and $d(\Delta)\hat{\sigma}^{\,W}_{i,ab}$ to CC processes with $V = W^{\pm}$.  The quantity $(\Delta)\overline{\Sigma}$ represents the sum (or difference, for polarized cases) over initial-state polarizations, 
and the average over final-state polarizations and color quantum numbers.  
The phase space for the final state, consisting of $X$ and $b$, is denoted by $\text{dPS}_{X+b}$.  In $d$ space-time dimensions, the projectors $(\Delta)\mathcal{P}^{\mu\nu}_i$ are written in terms of partonic tensors defined as $t_{1,\mu\nu} = T_{1,\mu\nu}$, $t_{2,\mu\nu}= x_1 T_{2,\mu\nu}$, and are given by
\begin{align}
\mathcal{P}^{\mu\nu}_1 =- \Delta\mathcal{P}^{\mu\nu}_5 &=\frac{1}{d-2}\left(t_{1,\mu\nu}+\frac{Q^2}{p_a\cdot q}t_{2,\mu\nu}\right) \,,\nonumber \\
\mathcal{P}^{\mu\nu}_2 = -\Delta \mathcal{P}^{\mu\nu}_4 &=\frac{Q^2}{(d-2)p_a.q}\left(t_{1, \mu\nu}+\frac{(d-1)Q^2}{p_a \cdot q}t_{2,\mu\nu}\right)\,, \nonumber \\
\frac{1}{2}\mathcal{P}^{\mu\nu}_3 = \Delta \mathcal{P}^{\mu\nu}_{1} &= \frac{i}{(d-2)(d-3)}
\epsilon^{\mu \nu \sigma \lambda} \,  \frac{q_\sigma p_{a,\lambda}}{p_a\cdot q}
\, .
\end{align}

\textit{Methodology:} 
The computation of the partonic cross sections follows the methodology described in our earlier works~\cite{Goyal:2023zdi,Goyal:2024emo,Goyal:2024tmo}. 
For completeness, we briefly summarize the main steps here. 
The Feynman diagrams are generated using \texttt{QGRAF}~\cite{Nogueira:1991ex}. 
The output of \texttt{QGRAF} is subsequently processed using a set of in-house routines written in \texttt{FORM}~\cite{Kuipers:2012rf,Ruijl:2017dtg}, which convert the diagrams into a suitable format for the application of QCD and electroweak Feynman rules. These routines are further used to perform the Dirac algebra, Lorentz contractions, and the simplification of color factors.
In the matrix element squared $|M_{ab}|^2_{j,\mu\nu}$, the Dirac matrix $\gamma_{5}$ arises due to the axial-vector coupling as well as from spin dependent (anit-)quark states.  In addition,  we encounter Levi-Civita tensor  from polarization of gluons and also from the projectors, see e.g.~\cite{Zijlstra:1993sh}.  Since both the $\gamma_5$ matrix and the Levi-Civita tensors are intrinsically four-dimensional objects, we need to choose a prescription to define them in $d=4+\varepsilon$ dimensions.
There are several schemes to do so, however none of them is known to preserve the chiral Ward identity.
In the present work, for the axial-vector coupling, we employ Larin's prescription \cite{Larin:1993tq} to define $\gamma_5$ in $d=4+\varepsilon$,   
\begin{align}\label{eq:laring5}
\gamma_{\mu}\gamma_5 = {\frac{i}{6}} ~\epsilon_{\mu \nu \sigma \lambda} \gamma^\nu \gamma^\sigma \gamma^\lambda.
\end{align}
and to project out the polarization of initial quarks we use, (see~\cite{Behring:2019tus,Schonwald:2019gmn,Bonino:2025bqa}), 
\begin{align}
\slash\!\!\!p_a\gamma_5 = {\frac{i}{2~p_a\cdot q}} ~\epsilon_{\mu \nu \sigma \lambda} \gamma^\mu\gamma^\nu p_a^\sigma q^\lambda \slash\!\!\!p_a
\, .
\end{align}
With the above identities, we have encountered a maximum of four Levi-Civita tensors which need to be contracted in a particular order see \cite{Moch:2015usa,Belusca-Maito:2023wah,Bonino:2025bqa}:
\begin{itemize}
    \item Those coming from the projector and the polarized parton are contracted first.
    \item Then we contract the tensors arising from the axial-vector couplings.
    \item Finally, the remaining Levi-Civita tensors (from polarized partons and the vertex) are contracted.   
\end{itemize}
The contraction is expressed in terms of a determinant containing metric tensors defined in $d = 4 + \varepsilon$ dimensions.
\begin{align}
  \label{eqn:LeviContract}
\epsilon_{\mu_{1}\nu_{1}\lambda_{1}\sigma_{1}} \, \epsilon_{\mu_{2}\nu_{2}\lambda_{2}\sigma_{2}} \!&=\! -
{\left | \!
  \begin{array}{cccc}
g_{\mu_{1} \mu_{2}}&g_{\mu_{1} \nu_{2}}&g_{\mu_{1} \lambda_{2}}&g_{\mu_{1} \sigma_{2}} \\
g_{\nu_{1} \mu_{2}}&g_{\nu_{1} \nu_{2}}&g_{\nu_{1} \lambda_{2}}&g_{\nu_{1} \sigma_{2}} \\
g_{\lambda_{1} \mu_{2}}&g_{\lambda_{1} \nu_{2}}&g_{\lambda_{1} \lambda_{2}}&g_{\lambda_{1} \sigma_{2}} \\
g_{\sigma_{1} \mu_{2}}&g_{\sigma_{1} \nu_{2}}&g_{\sigma_{1} \lambda_{2}}&g_{\sigma_{1} \sigma_{2}}
  \end{array} \!
\right |}
\, ,
\end{align}
such that in $d = 4$, $\epsilon_{\mu\nu\lambda\sigma}\epsilon^{\mu\nu\lambda\sigma} = -24$.
The results obtained in Larin's scheme are converted to the $\overline {\rm{MS}}$ scheme through a finite renormalization scheme transformation.  
The renormalization constants~\cite{Matiounine:1998re,Ravindran:2003gi,Moch:2014sna} used for this scheme transformation restore the Ward identity and define the CFs in ${\overline {\rm{MS}}}$ scheme. Further details will be presented below.

Beyond leading order (LO) in perturbation theory, the partonic cross sections $d(\Delta )\hat{\sigma}^j_{i,ab}$ can be categorized into three types: purely virtual (VV), purely real emissions (RR), and real-virtual (RV) contributions. The VV component includes one-loop and two-loop QCD corrections to the Born subprocess. The RV and RR contributions involve single-parton real emissions, with the RV terms containing one-loop corrections to these emissions, and the RR terms involving processes with both single and double real emissions.
The partonic cross sections $d (\Delta) \hat \sigma^j_{i,ab}$ contain both ultraviolet (UV) and infrared (IR) divergences. 
The UV divergences stem from virtual corrections to the partonic subprocesses, while the IR divergences originate from soft and collinear configurations that appear in both virtual and real emission subprocesses. We have used dimensional regularization with $d = 4+\varepsilon$ space-time dimensions to regulate both UV and IR divergences.

In the RV and RR contributions, due to the presence of the SIDIS constraint $\delta\left(\frac{z}{z_1} - \frac{p_a.p_b}{p_a.q}\right)$, the corresponding phase space integrals are more involved compared to fully inclusive DIS. 
With reverse unitarity~\cite{Anastasiou:2003gr,Anastasiou:2012kq}, we have mapped all phase space integrals to loop integrals and applied integration-by-parts identities (IBP)~\cite{Chetyrkin:1981qh,Laporta:2001dd}, to reduce them to master integrals (MIs). 
We have used \texttt{LiteRed}~\cite{Lee:2013mka} to generate the IBP rules for all VV, RV and RR contributions. 
The IBP reduction yields 4 MIs for VV, 7 MIs for RV, and 20 MIs for RR subprocesses. The results for VV MIs can be found in~\cite{Matsuura:1988sm, Matsuura:1987wt, Gehrmann:2005pd}, the MIs of RV are given in~\cite{Matsuura:1988sm, Gehrmann:2022cih, Goyal:2024emo}, and the MIs of RR can be found in~\cite{Bonino:2024adk, Ahmed:2024owh}.
 
At this stage all UV and IR divergences in the perturbative expansion of the partonic cross sections (computed in powers of the bare coupling $\hat a_s =\hat g_s^2/(16 \pi^2)$) appear as poles in $\varepsilon$. 
The UV divergences present in the loop corrections (VV and RV-terms) are removed by renormalization of the strong coupling in the $\rm{{\overline {MS}}}$ scheme:
\begin{eqnarray}
S_\varepsilon {\hat a_s \over (\mu^2)^{\varepsilon/2}}
={a_s(\mu_R^2) \over(\mu_R^2)^{\varepsilon/2}}
\left(1+a_s(\mu_R^2) {2 \beta_0 \over \varepsilon} + {\cal O}(a_s^2)\right)
\, ,
\end{eqnarray}
where $\mu$ is the 't Hooft scale, used to render $\hat a_s$ dimensionless, and $S_\varepsilon = \exp\Big({\varepsilon\over 2}(\gamma_E-\ln(4 \pi))\Big)$ is the spherical factor in $4+\varepsilon$ dimensions. $\beta_0={11\over 3} C_A-{4\over 3} T_f n_f$ is the LO coefficient of the QCD $\beta$ function with $C_A=N$ for an $SU(N)$ gauge group, $T_f=1/2$ and $n_f$ is number of active quarks. $\gamma_E$ is the Euler-Mascheroni constant.

Since we use Eq.~\eqref{eq:laring5} to define the $\gamma_5$ matrix, we need to use the $\overline{\text{MS}}$ scheme counter term $Z_{A5} = Z_{A}Z_{5}$~\cite{Larin:1993tq,Moch:2008fj} before performing the mass factorization. 
We expand $Z_{A5}$ in powers of $a_s$ as $Z_{A5} = 1 + \sum_{k=1} a_s^k(\mu_R^2)~Z_{A5}^{(k)}$, where
\begin{widetext}
\begin{align}
\label{ep:ZA5}
Z_{A5}^{(1)} &=  C_F \left(-4 +5 \varepsilon
  - \varepsilon^2\left(\frac{11}{2} - \frac{1}{2} \zeta_{2}\right)
\right)\, ,
\nonumber\\
Z_{A5}^{(2)} &=  C_F^2 \big(22\big) -C_A C_F \left( \frac{44}{3\varepsilon} +\frac{107}{9}\right)  +C_F n_f \left( \frac{8}{3\varepsilon} +\frac{2}{9} \right)\,. 
\end{align}    
\end{widetext}
The soft singularities from virtual and real emission subprocesses cancel among themselves.
Collinear singularities from initial- and final-state partons are removed via mass factorization: the space-like Altarelli-Parisi (AP) kernels, denoted by 
$\Gamma_{c\leftarrow a}$, renormalize the PDFs, while the time-like AP kernels, $\tilde{\Gamma}_{b\leftarrow d}$renormalize the FFs. This is performed at the scale $\mu_F$ and can be written as:
\begin{align}
\label{eq:massfact}
 d (\Delta)\hat \sigma^j_{i,ab}(\varepsilon)  =&  
 \Gamma_{c\leftarrow a}(\varepsilon,\mu_F^2) \otimes 
(\Delta){\cal C}^j_{i,cd}(\varepsilon,\mu_F^2) \nonumber \\
&\tilde{\otimes} 
\tilde{\Gamma}_{b\leftarrow d}(\varepsilon,\mu_F^2)
\, .
\end{align}
Here the summation over $c,d$ is implied. The symbols $\otimes$ and $\tilde{\otimes}$ denote convolutions over the scaling variables associated with PDFs and FFs, respectively, namely $x' = x/x_1$ and $z' = z/z_1$ (see Eq.~(\ref{eq:SFdef})).
For the unpolarized SFs, Eq.~\eqref{eq:massfact} yields the CFs in the $\overline{\rm{MS}}$ and the counter terms in Eq.~\eqref{ep:ZA5} completely remove the scheme dependence from Larin's prescription.  In the case of polarized SFs, these counter terms remove the scheme dependence from the $\gamma_5$ matrices that are present in the vertices. 
The treatment of the $\gamma_5$ matrix that enters through the polarization of incoming partons requires an additional finite renormalization to obtain the CFs in $\overline{\rm{MS}}$ scheme.  This is done in two steps. We first perform mass factorization for the polarized CFs using the space-like polarized AP kernels computed in Larin's  scheme and then perform a suitable scheme transformations to convert these Larin scheme CFs to
$\overline{\rm{MS}}$ ones.
Following~\cite{Moch:2014sna} the finite renormalization constants $Z_{ab}$~\cite{Matiounine:1998re,Ravindran:2003gi,Moch:2014sna} transform the CFs, defined in the Larin scheme, denoted by $\Delta {\cal C}^L_{i,ab}$ to those in the $\overline {\rm MS}$ scheme, $\Delta{\cal C}_{i,ab}$, 
\begin{equation}\label{eq:larinG}
 \Delta  {\cal C}_{1,ab}(\mu_F^2)= \big(Z^{-1}(\mu_F^2)\big)_{ad}\otimes \Delta {\cal C}_{1,db}^{L} (\mu_F^2)\,.
\end{equation}
The symmetric matrix valued $Z$-factor, e.g. given in \cite{Matiounine:1998re,Ravindran:2003gi,Moch:2014sna}, can be expanded in powers of $a_s$ as
\begin{align}
Z_{ab}(x) = \delta(1-x) + \sum_{k=1}^{\infty} a_s^{k} z_{ab}^{(k)}(x) \,.
\end{align}
Using $z_{qg}^{(k)} =z_{\overline{q}g}^{(k)} 
=z_{gg}^{(k)} =0$ along with

\begin{align}
Z_{q_{i}q_{j}} &= \delta(1-x) + a_s z^{(1)}_{q_{i}q_{j}}+a_s^2 z^{(2)}_{q_{i}q_{j}} +{\mathcal{O}}(a_s^3)\ ,\nonumber \\[4pt]
Z_{q_{i}\overline{q}_{j}} &= \delta(1-x) + a_s z^{(1)}_{q_{i}\overline{q}_{j}}+a_s^2 z^{(2)}_{q_{i}\overline{q}_{j}} +{\mathcal{O}}(a_s^3)\, ,
\end{align}

%
where $i, j$ denote quarks flavors, the $\overline{\rm{MS}}$ scheme CFs read,
\begin{align}
\label{eq:CFs-1}
\Delta {\cal{C}}^{(0)}_{i,qq} &=\Delta {\cal{C}}^{L,(0)}_{i,qq} 
\, ,\nonumber\\
\Delta{\cal{C}}^{(1)}_{i,qq} &=\Delta {\cal{C}}^{L,(1)}_{i,qq} + (-z^{(1)}_{qq})\otimes \Delta {\cal{C}}^{L,(0)}_{i,qq} \, ,\nonumber\\
\Delta{\cal{C}}^{(1)}_{i,qg} &= \Delta{\cal{C}}^{L,(1)}_{i,qg} \, ,\nonumber\\
\Delta{\cal{C}}^{(1)}_{i,gq} &= \Delta{\cal{C}}^{L,(1)}_{i,gq} \, ,\nonumber\\
\Delta{\cal{C}}^{(2)}_{i,qq} &= \Delta{\cal{C}}^{L,(2)}_{i,qq} + (-z^{(1)}_{qq})\otimes \Delta{\cal{C}}^{L,(1)}_{i,qq}  \nonumber\\&+ \Big( z^{(1)}_{qq} \otimes z^{(1)}_{qq}  -  z^{(2)}_{qq}  \Big) \otimes \Delta{\cal{C}}^{L,(0)}_{i,qq} \, ,\nonumber\\
\Delta{\cal{C}}^{(2)}_{i,q\overline{q}}  &=  \Delta{\cal{C}}^{L,(2)}_{i,q\overline{q}} + \Big(-  z^{(2)}_{q\overline{q}}  \Big) \otimes \Delta{\cal{C}}^{L,(0)}_{i,\overline{q}\overline{q}} \, ,\nonumber\\
 \Delta{\cal{C}}^{(2)}_{i,qq'}  &=  \Delta{\cal{C}}^{L,(2)}_{i,qq'} + \Big(-  z^{(2)}_{qq'} \Big) \otimes \Delta{\cal{C}}^{L,(0)}_{i,q'q'} \, ,\nonumber\\
\Delta{\cal{C}}^{(2)}_{i,q\overline{q}'}  &=  \Delta{\cal{C}}^{L,(2)}_{i,q\overline{q}'} + \Big(-  z^{(2)}_{qq'} \Big) \otimes \Delta{\cal{C}}^{L,(0)}_{i,\overline{q}'\overline{q}'} \, ,\nonumber\\
\Delta{\cal{C}}^{(2)}_{i,qg} &= \Delta{\cal{C}}^{L,(2)}_{i,qg} + (-z^{(1)}_{qq})\otimes \Delta{\cal{C}}^{L,(1)}_{i,qg} \, ,\nonumber\\
\Delta{\cal{C}}^{(2)}_{i,gq} &= \Delta{\cal{C}}^{L,(2)}_{i,gq}\, ,
\hspace{0.3cm} \Delta{\cal{C}}^{(2)}_{i,gg} = \Delta{\cal{C}}^{L,(2)}_{i,gg}
\, .
\end{align}
The expressions for  $z_{ab}^{(k)}$ are given in \cite{Goyal:2024emo}. 
For completeness we present them here,
\begin{align}
z_{q_{i}q_{j}}^{(k)} &= z_{\overline{q}_{i}\overline{q}_{j}}^{(k)} = \delta_{ij}z_{qq}^{(k),\rm V} + z_{qq}^{(k),\rm S}\, ,\nonumber\\
z_{q_{i}\overline{q}_{j}}^{(k)} &= z_{\overline{q}_{i}{q}_{j}}^{(k)} =\delta_{ij}z_{q\overline{q}}^{(k),\rm V} + z_{q\overline{q}}^{(k),\rm S}\, ,
\end{align}
where the $z^{(k)}_{ab}$ coefficients are given by:
\begin{widetext}
\begin{align}
z_{q_{i}q_{j}}^{(1)}(x) &= \delta_{ij} C_F \Big\{
       - 8 + 8 x\Big\}\, , ~~~~ z_{{q}_{i}\overline{q}_{j}}^{(1)}(x) = 0 \,,\nonumber \\   
z_{qq}^{(2),\rm V}(x) &=   C_F n_f T_F \bigg\{
        \frac{16}{3}\big( 1 - x \big)\ln(x)  
       + \frac{80}{9} \big( 1- x \big)\bigg\}
 -
   C_A C_F \bigg\{
           \bigg(   \frac{80}{3} - \frac{8}{3} x \bigg)\ln(x)
       + 4   \big(   1 -  x \big)\ln^2(x)
\nonumber\\&       
       + \frac{592}{9} \big(1-x\big)  - 8 \zeta_2 \big(1-x\big)\bigg\}
 +
   C_F^2 \bigg\{
        16   \big( 1- x \big)\ln(x) \ln(1-x) 
       - 8   \big(   2 +  x \big)\ln(x)       
       - 16\big(1-x\big) \bigg\}\,,\nonumber \\
z_{q\overline{q}}^{(2),\rm V}(x) &= -\bigg(C_F^2 - \frac{1} {2}C_A C_F\bigg) \bigg\{ 8 \big(1+x\big) \Big(
        4 \text{Li}_{2}(-x)
       + 4 \ln(x) \ln(1+x)  
       - \ln^2(x)           
       - 3 \ln(x) 
       + 2 \zeta_2 \Big)
       - 56 \big( 1 - x \big) \bigg\}
\,,\nonumber \\
z_{qq}^{(2),\rm S}(x) &= z_{q\overline{q}}^{(2),\rm S}(x) = C_FT_F \bigg\{
        8 \big( 3 -  x \big) \ln(x)  
       +4 \big( 2 +  x \big) \ln^2(x)   
       + 16 \big( 1 - x \big) \bigg\}\, .
\end{align}
%
After obtaining the CFs for both unpolarized and polarized cases, we substitute them into Eq.~\eqref{eq:SFdef} to derive the SFs. 
Next, we present the SFs expressed as convolutions of CFs, PDFs, and FFs for NC and CC SIDIS processes, seeting $\mu_F^2=\mu_F^2=Q^2$, 
and expandeding as
\begin{align}
{\cal F}_i^j &= \sum_{k=0}^{\infty} a_s^k (\mu_R^2) {\cal F}_i^{j,(k)}(\mu_R^2)
\, .
\end{align}
In case of $\gamma$ exchange, the SFs  ${\cal F}^{\gamma\gamma}_i\,, i=1,2$ up to $a_s^2$ read
%
\begin{align}
\label{eq:nnlo-FggI}
{\cal F}_{i}^{\gamma\gamma,(0)} = &\sum_{q} e_q^{2}~H_{qq}\hat\otimes C_{I,qq}^{\gamma\gamma,(0)}\,,\\
{\cal F}^{\gamma\gamma,(1)}_{i}  =& \sum_{q}  e_q^{2} \bigg(
H_{qq}\hat \otimes C_{I,qq}^{\gamma\gamma,(1)} + 
H_{qg} \hat \otimes C_{I,qg}^{\gamma\gamma,(1)} + 
H_{gq} \hat \otimes C_{I,gq}^{\gamma\gamma,(1)} \bigg) 
\, ,\\
{\cal F}^{\gamma\gamma,(2)}_i = &
\sum_{q} e_q^{2} ~H_{qq}  \hat \otimes  C_{I,qq}^{\gamma\gamma,(2),\rm NS} +
\sum_{q,q_{k}} e_{q_k}^{2} \bigg(H_{qq}\hat \otimes  C_{I,qq}^{\gamma\gamma,(2),\rm PS} +  
H_{gg} \hat \otimes C_{I,gg}^{\gamma\gamma,(2)}\bigg)\nonumber\\
& + \sum_{q} e_q^{2} \bigg( H_{qg} \hat \otimes C_{I,qg}^{\gamma\gamma,(2)} +
H_{gq} \hat \otimes C_{I,gq}^{\gamma\gamma,(2)} 
+
H_{q\bar{q}} \hat \otimes C_{I,q\bar{q}}^{\gamma\gamma,(2)} 
\bigg)
\nonumber\\
&+
\sum_{q} \sum_{q'\neq q}\bigg( e_q^{2}~H^{+}_{qq'} \hat \otimes  C_{I,qq'}^{\gamma\gamma,(2),[1]} +
e_{q'}^{2} ~H^{+}_{qq'}\hat \otimes  C_{I,qq'}^{\gamma\gamma,(2),[2]} + 
e_q e_{q'} ~ H^{-}_{qq'} \hat \otimes C_{I,qq'}^{\gamma\gamma,(2),[3]}\bigg)
\, .
\end{align}
For $\gamma Z$ interference, the SFs  ${\cal F}^{\gamma Z}_i\,, i=1,2$ up to $a_s^2$ are given by
%
\begin{align}
\label{eq:nnlo-FgZI}
-{\cal F}_{i}^{\gamma Z,(0)} = &\sum_{q} 2e_{q}\overline{g}_{V_q}~ H_{qq}\hat\otimes C_{I,qq}^{\gamma\gamma,(0)}\,,\\
-{\cal F}^{\gamma Z,(1)}_{i}  =& \sum_{q} 2e_{q}\overline{g}_{V_q}\bigg(
H_{qq}\hat \otimes C_{I,qq}^{\gamma\gamma,(1)} + 
H_{qg} \hat \otimes C_{I,qg}^{\gamma\gamma,(1)} + 
H_{gq} \hat \otimes C_{I,gq}^{\gamma\gamma,(1)} \bigg) 
\, ,\\
-{\cal F}^{\gamma Z,(2)}_I = &
\sum_{q} 2e_{q}\overline{g}_{V_q}~H_{qq}  \hat \otimes  C_{I,qq}^{\gamma\gamma,(2),\rm NS} +
\sum_{q,q_{k}} 2e_{q_k}\overline{g}_{V_{q_k}} \bigg(H_{qq}\hat \otimes  C_{I,qq}^{\gamma\gamma,(2),\rm PS} +  
H_{gg} \hat \otimes C_{I,gg}^{\gamma\gamma,(2)}\bigg)\nonumber\\
& + \sum_{q} 2 e_{q}\overline{g}_{V_q} \bigg( H_{qg} \hat \otimes C_{I,qg}^{\gamma\gamma,(2)} +
H_{gq} \hat \otimes C_{I,gq}^{\gamma\gamma,(2)} 
+
H_{q\bar{q}} \hat \otimes C_{I,q\bar{q}}^{\gamma\gamma,(2)} \bigg) 
\nonumber\\
&+
\sum_{q} \sum_{q'\neq q}\bigg( 2 e_{q}\overline{g}_{V_q}~H^{+}_{qq'} \hat \otimes  C_{I,qq'}^{\gamma\gamma,(2),[1]} +
2e_{q'}\overline{g}_{V_{q'}} ~H^{+}_{qq'}\hat \otimes  C_{I,qq'}^{\gamma\gamma,(2),[2]} + 
\left(e_{q'}\overline{g}_{V_q} + e_{q}\overline{g}_{V_{q'}}\right) H^{-}_{qq'} \hat \otimes C_{I,qq'}^{\gamma\gamma,(2),[3]}
\bigg)
\, ,
\end{align}
%
and for the SF ${\cal F}^{\gamma Z}_3$ we find,

\begin{align}
\label{eq:nnlo-FgZ3}
{\cal F}_{3}^{\gamma Z,(0)} = &\sum_{q} 2e_{q} \overline{g}_{A_q}~ H_{qq}\hat\otimes C_{3,qq}^{(0)}\,,\\
{\cal F}^{\gamma Z,(1)}_{3}  =& \sum_{q} 2e_{q} \overline{g}_{A_q} \bigg(
H_{qq}\hat \otimes C_{3,qq}^{(1)} + 
H_{qg} \hat \otimes C_{3,qg}^{(1)} + 
H_{gq} \hat \otimes C_{3,gq}^{(1)} \bigg) 
\, ,\\
{\cal F}^{\gamma Z,(2)}_3 = &
 \sum_{q} 2e_{q} \overline{g}_{A_q} \bigg(
H_{qq}  \hat \otimes C_{3,qq}^{(2),\rm NS} +  
H_{qg} \hat \otimes C_{3,qg}^{(2)} +
H_{gq} \hat \otimes C_{3,gq}^{(2)} 
+
H_{q\bar{q}} \hat \otimes  C_{3,q\bar{q}}^{(2)} \bigg)
\nonumber\\
&+
\sum_{q} \sum_{q'\neq q}\bigg( 
2e_{q} \overline{g}_{A_q}~H^{+}_{qq'} \hat \otimes  C_{3,qq'}^{(2),[1]} +
 2e_{q'} \overline{g}_{A_{q'}}  ~H^{-}_{qq'}\hat \otimes  C_{3,qq'}^{(2),[2]} + 
2e_{q} \overline{g}_{A_{q'}}~ H^{+}_{qq'} \hat \otimes C_{3,qq'}^{(2),[3]} 
+ 2e_{q'} \overline{g}_{A_q}~ H^{-}_{qq'} \hat \otimes C_{3,qq'}^{(2),[4]}
\bigg)
\, .
\end{align} 
In case of $Z$-boson exchange, the SFs  ${\cal F}^{ZZ}_i\,, i=1,2$ up to $a_s^2$ we have
%
\begin{align}
\label{eq:nnlo-FZZI}
{\cal F}_{i}^{ZZ,(0)} = &\sum_{q} \left(\overline{g}_{V_q}^{2}+\overline{g}_{A_q}^{2}\right) H_{qq}\hat\otimes C_{I,qq}^{\gamma\gamma,(0)}\,,\\
{\cal F}^{ZZ,(1)}_{i}  =& \sum_{q} \left(\overline{g}_{V_q}^{2}+\overline{g}_{A_q}^{2}\right)\bigg(
H_{qq}\hat \otimes C_{I,qq}^{\gamma\gamma,(1)} + 
H_{qg} \hat \otimes C_{I,qg}^{\gamma\gamma,(1)} + 
H_{gq} \hat \otimes C_{I,gq}^{\gamma\gamma,(1)} \bigg) 
\, ,\\
{\cal F}^{ZZ,(2)}_i = &
\sum_{q}  H_{qq}  \hat \otimes \bigg(\overline{g}_{V_q}^{2} C_{I,qq}^{\gamma\gamma,(2),\rm NS} +
\overline{g}_{A_q}^{2}  C_{I,qq}^{AA,(2),\rm NS} \bigg) +
\sum_{q,q_{k}} \left(\overline{g}_{V_{q_{k}}}^{2}+\overline{g}_{A_{q_{k}}}^{2}\right) \bigg(H_{qq}\hat \otimes  C_{I,qq}^{\gamma\gamma,(2),\rm PS} +  
H_{gg} \hat \otimes C_{I,gg}^{\gamma\gamma,(2)}\bigg)\nonumber\\
& + \sum_{q} \left(\overline{g}_{V_q}^{2}+\overline{g}_{A_q}^{2}\right) \bigg( H_{qg} \hat \otimes C_{I,qg}^{\gamma\gamma,(2)} +
H_{gq} \hat \otimes C_{I,gq}^{\gamma\gamma,(2)} \bigg)
+
\sum_{q} H_{q\bar{q}} \hat \otimes \bigg(\overline{g}_{V_q}^{2} C_{I,q\bar{q}}^{\gamma\gamma,(2)} +
\overline{g}_{A_q}^{2} C_{I,q\bar{q}}^{AA,(2)} \bigg)
\nonumber\\
&+
\sum_{q} \sum_{q'\neq q}\bigg( \left(\overline{g}_{V_q}^{2}+\overline{g}_{A_q}^{2}\right)~H^{+}_{qq'} \hat \otimes  C_{I,qq'}^{\gamma\gamma,(2),[1]} +
\left(\overline{g}_{V_{q'}}^{2} + \overline{g}_{A_{q'}}^{2}\right)  ~H^{+}_{qq'}\hat \otimes  C_{I,qq'}^{\gamma\gamma,(2),[2]} + 
\overline{g}_{V_q} \overline{g}_{V_{q'}} H^{-}_{qq'} \hat \otimes C_{I,qq'}^{\gamma\gamma,(2),[3]} \nonumber\\
& + \overline{g}_{A_q} \overline{g}_{A_{q'}} H^{+}_{qq'} \hat \otimes C_{I,qq'}^{AA,(2),[3]}
\bigg)
\, ,
\end{align}
%
and for the SF ${\cal F}^{ZZ}_3$ we find
%
\begin{align}
\label{eq:nnlo-FZZ3}
-{\cal F}_{3}^{ZZ,(0)} = &\sum_{q} \left(2\overline{g}_{V_q}\overline{g}_{A_q}\right) H_{qq}\hat\otimes C_{3,qq}^{(0)}\,,\\
-{\cal F}^{ZZ,(1)}_{3}  =& \sum_{q} \left(2 \overline{g}_{V_q}\overline{g}_{A_q} \right)\bigg(
H_{qq}\hat \otimes C_{3,qq}^{(1)} + 
H_{qg} \hat \otimes C_{3,qg}^{(1)} + 
H_{gq} \hat \otimes C_{3,gq}^{(1)} \bigg) 
\, ,\\
-{\cal F}^{ZZ,(2)}_3 = &
 \sum_{q} \left(2\overline{g}_{V_q}\overline{g}_{A_q}\right) \bigg(
H_{qq}  \hat \otimes C_{3,qq}^{(2),\rm NS} +  
H_{qg} \hat \otimes C_{3,qg}^{(2)} +
H_{gq} \hat \otimes C_{3,gq}^{(2)} 
+
H_{q\bar{q}} \hat \otimes  C_{3,q\bar{q}}^{(2)} \bigg)
\nonumber\\
&+
\sum_{q} \sum_{q'\neq q}\bigg( 
\left(2\overline{g}_{V_q}\overline{g}_{A_q}\right)~H^{+}_{qq'} \hat \otimes  C_{3,qq'}^{(2),[1]} +
\left(2\overline{g}_{V_{q'}}\overline{g}_{A_{q'}}\right)  ~H^{-}_{qq'}\hat \otimes  C_{3,qq'}^{(2),[2]} + 
\left(2\overline{g}_{V_q} \overline{g}_{A_{q'}}\right) H^{+}_{qq'} \hat \otimes C_{3,qq'}^{(2),[3]} \nonumber\\
& + \left(2\overline{g}_{V_{q'}} \overline{g}_{A_{q}}\right) H^{-}_{qq'} \hat \otimes C_{3,qq'}^{(2),[4]}
\bigg)
\, ,
\end{align} 
where
\begin{align}
H_{qq} &=  f_q(x) D_q(z) + \eta~ f_{\bar{q}} (x)D_{\bar{q}}(z)\, , 
\quad
H_{q\bar{q}} = f_q (x) D_{\bar{q}} (z) +\eta~f_{\bar{q}}(x) D_q(z)\, , \nonumber\\
H_{qg} &= f_q (x) D_g (z) +\eta~ f_{\bar{q}}(x) D_g (z)\, , 
\quad
H_{gq} = f_g (x) D_q (z) +\eta~ f_g D_{\bar{q}}(z)\, ,\nonumber\\  
H^{\pm}_{qq'} &= f_q (x)D_{q'}(z) \pm f_q (x)D_{\bar{q}'}(z) 
\pm \eta~ f_{\bar{q}} (x)D_{q'}(z) +\eta~ f_{\bar{q}} (x) D_{\bar{q}'}(z)\, ,\nonumber\\
H_{gg} &= f_g (x)D_g(z)\, .
\end{align}
Here, we have $\eta=1$ for ${\cal F}_{i}^{j,(k)}$ and $\eta=-1$ for ${\cal F}^{j,(k)}_3$ for $j = \gamma\gamma,\gamma Z,ZZ$. 
Also, $\overline{g}_{V_{f}}$  and $\overline{g}_{A_{f}}$ are defined in Eqs.~\eqref{eq:gAgVbar}, \eqref{eq:gAgV}.
For CC SIDIS, the SFs ${\cal F}_i^{W^-}, i=1,2$ are found to be
%
\begin{align}
\label{eq:lo-Wm}
\frac{1}{2} {\cal F}_{i}^{W^{-},(0)}&=\sum_{\um,\bm} \big|\overline{V}_{\um\bm}\big|^{2}\left(
H_{\um\bm} +  H_{\overline{\bm}\overline{\um}}\right)\hat{\otimes}\text{C}^{\gamma\gamma,(0)}_{I,qq}\, ,
\\
\label{eq:nlo-Wm}
\frac{1}{2}{\cal F}_{i}^{W^{-},(1)}& =\! \sum_{\um,\bm} \big|\overline{V}_{\um\bm}\big|^{2}\bigg(
\Big( H_{\um\bm} +  H_{\overline{\bm}\overline{\um}}\Big) \hat{\otimes}\text{C}^{\gamma\gamma, (1)}_{I,qq}
+\Big( H_{\um g} + H_{\overline{\bm} g} \Big)\hat{\otimes} \text{C}^{\gamma\gamma,(1)}_{I,q g}
+ \Big( H_{g\bm} 
+ H_{g\overline{\um}} \Big)\hat{\otimes} \text{C}^{\gamma\gamma,(1)}_{I,g q}  \bigg)\, , 
\\
\label{eq:nnlo-Wm}
\frac{1}{2}{\cal F}_{i}^{W^{-},(2)}& =
\sum_{\um,\bm} \big|\overline{V}_{\um\bm}\big|^{2}\left( H_{\um\bm} +  H_{\overline{\bm}\overline{\um}}\right) \hat{\otimes}\text{C}^{W^-, (2),[1]}_{I,q q}
+ \sum_{\um,\bm} \bigg( \bigg(\sum_{\beta}\big|\overline{V}_{\um\beta }\big|^{2}\bigg) H_{\um \bm} + \bigg(\sum_{\alpha}\big|\overline{V}_{\alpha\bm }\big|^{2}\bigg) H_{\overline{\bm}\overline{\um} }\bigg) \hat{\otimes} \text{C}^{\gamma\gamma,(2),[1]}_{I,qq'}
\nonumber\\&
+ \sum_{\um,\bm} \bigg( \bigg(\sum_{\alpha}\big|\overline{V}_{\alpha \bm }\big|^{2}\bigg) H_{\um \bm} + \bigg(\sum_{\beta}\big|\overline{V}_{\um\beta }\big|^{2}\bigg) H_{\overline{\bm}\overline{\um} }\bigg) \hat{\otimes} \text{C}^{\gamma\gamma,(2),[2]}_{I,qq'} 
+ \sum_{\um,\bm}\big|\overline{V}_{\um \bm }\big|^{2}\left( H_{\um g} +  H_{\overline{\bm} g }\right) \hat{\otimes} \text{C}^{\gamma\gamma,(2)}_{I,q g} 
\nonumber\\&
 + \sum_{\um,\bm}\big|\overline{V}_{\um \bm }\big|^{2}\left( H_{\um \um} + H_{\overline{\bm} \overline{\bm} }\right) \hat{\otimes} \text{C}^{W^-,(2),[2]}_{I,qq} + 
\bigg(\sum_{\alpha,\beta}\big|\overline{V}_{\alpha \beta }\big|^{2}\bigg)\left(  \sum_{\um}  H_{\um \um} + \sum_{\bm} H_{\overline{\bm} \overline{\bm} }\right) \hat{\otimes} \text{C}^{\gamma\gamma,(2),\text{PS}}_{I,q q}
\nonumber\\&
 + \sum_{\um,\bm}\big|\overline{V}_{\um \bm }\big|^{2}\left( H_{\bm \bm} + H_{\overline{\um} \overline{\um} }\right) \hat{\otimes} \text{C}^{W^-,(2),[3]}_{I,qq} + 
\bigg(\sum_{\alpha,\beta}\big|\overline{V}_{\alpha \beta }\big|^{2}\bigg)\left(  \sum_{\bm}  H_{\bm \bm} + \sum_{\um} H_{\overline{\um} \overline{\um} }\right) \hat{\otimes} \text{C}^{\gamma\gamma,(2),\text{PS}}_{I,q q}
\nonumber\\&
+ \sum_{\um,\bm}\big|\overline{V}_{\um \bm }\big|^{2}\left( H_{\um \overline{\um}} + H_{\overline{\bm} \bm }\right) \hat{\otimes} \text{C}^{W^-,(2),[1]}_{I,q \overline{q}}
+ \sum_{\um,\bm}\bigg( \bigg(\sum_{\beta}\big|\overline{V}_{\um \beta }\big|^{2}\bigg) H_{\um \overline{\bm}} + \bigg(\sum_{\alpha}\big|\overline{V}_{\alpha \bm }\big|^{2}\bigg) H_{\overline{\bm} \um}\bigg) \hat{\otimes} \text{C}^{\gamma\gamma,(2),[1]}_{I,qq'}
\nonumber\\&
+ \sum_{\um,\bm}\big|\overline{V}_{\um \bm }\big|^{2}\left( H_{\um \overline{\bm}} + H_{\overline{\bm} \um}\right) \hat{\otimes} \text{C}^{W^-,(2),[2]}_{I,q\overline{q}} 
+ \sum_{\um,\bm}\bigg( \bigg(\sum_{\beta}\big|\overline{V}_{\um \beta }\big|^{2}\bigg) H_{\bm \overline{\um}} + \bigg(\sum_{\alpha}\big|\overline{V}_{\alpha \bm }\big|^{2}\bigg) H_{\overline{\um} \bm}\bigg) \hat{\otimes} \text{C}^{\gamma\gamma,(2),[2]}_{I,qq'}
\nonumber\\&
+ \sum_{\um,\bm}\big|\overline{V}_{\um \bm }\big|^{2}\left( H_{\bm \overline{\um}} + H_{\overline{\um} \bm}\right) \hat{\otimes} \text{C}^{W^-,(2),[3]}_{I,q\overline{q}} 
+\bigg( \sum_{\um,\um'}\bigg(\sum_{\beta}\big|\overline{V}_{\um \beta }\big|^{2}\bigg) H_{\um \um'} + \sum_{\bm,\bm'} \bigg(\sum_{\alpha}\big|\overline{V}_{\alpha \bm }\big|^{2}\bigg)H_{\overline{\bm} \overline{\bm}'}\bigg) \hat{\otimes} \text{C}^{\gamma\gamma,(2),[1]}_{I,qq'}
\nonumber\\&
+\bigg(\sum_{\bm,\bm'} \bigg(\sum_{\alpha} \big|\overline{V}_{\alpha \bm' }\big|^{2} \bigg)H_{\bm\bm'} +
\sum_{\um,\um'} \bigg(\sum_{\beta}\big|\overline{V}_{\um' \beta }\big|^{2}\bigg) H_{\overline{\um} \overline{\um}'}
\bigg) \hat{\otimes} \text{C}^{\gamma\gamma,(2),[2]}_{I,q q'}
+ \sum_{\um,\bm} \big|\overline{V}_{\um\bm }\big|^{2} 
\Big( H_{g\bm} + H_{g\overline{\um}} \Big)\hat{\otimes} \text{C}^{\gamma\gamma,(2)}_{I,g q}  
\nonumber\\&
+  \sum_{\um,\bm} \big|\overline{V}_{\um\bm }\big|^{2} 
 H_{gg} \hat{\otimes} \text{C}^{\gamma\gamma,(2)}_{I,g g} 
+\bigg( \sum_{\um,\um'}\bigg(\sum_{\beta}\big|\overline{V}_{\um \beta }\big|^{2}\bigg) H_{\um \overline{\um}'} + \sum_{\bm,\bm'} \bigg(\sum_{\alpha}\big|\overline{V}_{\alpha \bm }\big|^{2}\bigg)H_{\overline{\bm} \bm'}\bigg) \hat{\otimes} \text{C}^{\gamma\gamma,(2),[1]}_{I,qq'}
\nonumber\\&
+\bigg( \sum_{\um,\um'}\bigg(\sum_{\beta}\big|\overline{V}_{\um' \beta }\big|^{2}\bigg) H_{\um \overline{\um}'} + \sum_{\bm,\bm'} \bigg(\sum_{\alpha}\big|\overline{V}_{\alpha \bm' }\big|^{2}\bigg)H_{\overline{\bm} \bm'}\bigg) \hat{\otimes} \text{C}^{\gamma\gamma,(2),[2]}_{I,qq'}\, .
\end{align}
%
For the SF ${\cal F}_3^{W^-}$, we find
\begin{align}
\label{eq:lo-Wm3}
-\frac{1}{2}{\cal F}_{3}^{W^{-},(0)}=&\sum_{\um,\bm} \big|\overline{V}_{\um\bm}\big|^{2}\left(
H_{\um\bm} -  H_{\overline{\bm}\overline{\um}}\right)\hat{\otimes}\text{C}^{(0)}_{3,qq}\, ,\\
%
\label{eq:nlo-Wm3}
-\frac{1}{2}{\cal F}_{3}^{W^{-},(1)} =& \sum_{\um,\bm} \big|\overline{V}_{\um\bm}\big|^{2}\bigg(
\Big( H_{\um\bm} -  H_{\overline{\bm}\overline{\um}}\Big) \hat{\otimes}\text{C}^{ (1)}_{3,qq}
+\Big( H_{\um g} - H_{\overline{\bm} g} \Big)\hat{\otimes} \text{C}^{(1)}_{3,q g}
+ \Big( H_{g\bm} 
- H_{g\overline{\um}} \Big)\hat{\otimes} \text{C}^{(1)}_{3,g q}  \bigg)\, , 
\\
\label{eq:nnlo-Wm3}
-\frac{1}{2}{\cal F}_{3}^{W^{-},(2)} =&
\sum_{\um,\bm} \big|\overline{V}_{\um\bm}\big|^{2}\left( H_{\um\bm} -  H_{\overline{\bm}\overline{\um}}\right) \hat{\otimes}\text{C}^{W^-, (2),[1]}_{3,q q}
+ \sum_{\um,\bm} \bigg( \bigg(\sum_{\beta}\big|\overline{V}_{\um\beta }\big|^{2}\bigg) H_{\um \bm} - \bigg(\sum_{\alpha}\big|\overline{V}_{\alpha\bm }\big|^{2}\bigg) H_{\overline{\bm}\overline{\um} }\bigg) \hat{\otimes} \text{C}^{(2),[1]}_{3,qq'}
\nonumber\\&
+ \sum_{\um,\bm} \bigg( \bigg(\sum_{\alpha}\big|\overline{V}_{\alpha \bm }\big|^{2}\bigg) H_{\um \bm} - \bigg(\sum_{\beta}\big|\overline{V}_{\um\beta }\big|^{2}\bigg) H_{\overline{\bm}\overline{\um} }\bigg) \hat{\otimes} \text{C}^{(2),[2]}_{3,qq'} 
+ \sum_{\um,\bm}\big|\overline{V}_{\um \bm }\big|^{2}\left( H_{\um g} -  H_{\overline{\bm} g }\right) \hat{\otimes} \text{C}^{(2)}_{3,q g} 
\nonumber\\&
+ \sum_{\um,\bm}\big|\overline{V}_{\um \bm }\big|^{2}\left( H_{\um \overline{\um}} - H_{\overline{\bm} \bm }\right) \hat{\otimes} \text{C}^{W^-,(2),[1]}_{3,q \overline{q}}
+ \sum_{\um,\bm}\bigg( \bigg(\sum_{\beta}\big|\overline{V}_{\um \beta }\big|^{2}\bigg) H_{\um \overline{\bm}} - \bigg(\sum_{\alpha}\big|\overline{V}_{\alpha \bm }\big|^{2}\bigg) H_{\overline{\bm} \um}\bigg) \hat{\otimes} \text{C}^{(2),[1]}_{3,qq'}
\nonumber\\&
+ \sum_{\um,\bm}\big|\overline{V}_{\um \bm }\big|^{2}\left( H_{\um \overline{\bm}} - H_{\overline{\bm} \um}\right) \hat{\otimes} \text{C}^{W^-,(2),[2]}_{3,q\overline{q}} 
+ \sum_{\um,\bm}\bigg(- \bigg(\sum_{\beta}\big|\overline{V}_{\um \beta }\big|^{2}\bigg) H_{\bm \overline{\um}} + \bigg(\sum_{\alpha}\big|\overline{V}_{\alpha \bm }\big|^{2}\bigg) H_{\overline{\um} \bm}\bigg) \hat{\otimes} \text{C}^{(2),[2]}_{3,qq'}
\nonumber\\&
+ \sum_{\um,\bm}\big|\overline{V}_{\um \bm }\big|^{2}\left( H_{\bm \overline{\um}} - H_{\overline{\um} \bm}\right) \hat{\otimes} \text{C}^{W^-,(2),[3]}_{3,q\overline{q}} 
+\bigg( \sum_{\um,\um'}\bigg(\sum_{\beta}\big|\overline{V}_{\um \beta }\big|^{2}\bigg) H_{\um \um'} - \sum_{\bm,\bm'} \bigg(\sum_{\alpha}\big|\overline{V}_{\alpha \bm }\big|^{2}\bigg)H_{\overline{\bm} \overline{\bm}'}\bigg) \hat{\otimes} \text{C}^{(2),[1]}_{3,qq'}
\nonumber\\&
+\bigg(\sum_{\bm,\bm'} \bigg(\sum_{\alpha} \big|\overline{V}_{\alpha \bm' }\big|^{2} \bigg)H_{\bm\bm'} -
\sum_{\um,\um'} \bigg(\sum_{\beta}\big|\overline{V}_{\um' \beta }\big|^{2}\bigg) H_{\overline{\um} \overline{\um}'}
\bigg) \hat{\otimes} \text{C}^{(2),[2]}_{3,q q'}
 + \sum_{\um,\bm}\big|\overline{V}_{\um \bm }\big|^{2}\left( H_{\um \um} - H_{\overline{\bm} \overline{\bm} }\right) \hat{\otimes} \text{C}^{W^-,(2),[2]}_{3,qq} 
\nonumber\\&
 + \sum_{\um,\bm}\big|\overline{V}_{\um \bm }\big|^{2}\left( H_{\bm \bm} - H_{\overline{\um} \overline{\um} }\right) \hat{\otimes} \text{C}^{W^-,(2),[3]}_{3,qq} 
+\bigg( \sum_{\um,\um'}\bigg(\sum_{\beta}\big|\overline{V}_{\um \beta }\big|^{2}\bigg) H_{\um \overline{\um}'} - \sum_{\bm,\bm'} \bigg(\sum_{\alpha}\big|\overline{V}_{\alpha \bm }\big|^{2}\bigg)H_{\overline{\bm} \bm'}\bigg) \hat{\otimes} \text{C}^{(2),[1]}_{3,qq'}
\nonumber\\&
+\bigg( -\sum_{\um,\um'}\bigg(\sum_{\beta}\big|\overline{V}_{\um' \beta }\big|^{2}\bigg) H_{\um \overline{\um}'} + \sum_{\bm,\bm'} \bigg(\sum_{\alpha}\big|\overline{V}_{\alpha \bm' }\big|^{2}\bigg)H_{\overline{\bm} \bm'}\bigg) \hat{\otimes} \text{C}^{(2),[2]}_{3,qq'}
+ \sum_{\um,\bm} \big|\overline{V}_{\um\bm }\big|^{2} 
\Big( H_{g\bm} - H_{g\overline{\um}} \Big)\hat{\otimes} \text{C}^{(2)}_{3,g q}  \, ,
\end{align}
\end{widetext}
with up-and down-type quarks, $\um,\alpha = \{u,c,(t)\}$ and $\bm, \beta = \{d,s,(b)\}$.
The symbol $\hat\otimes$ denotes the convolutions in both variables $x$ and $z$ and $H_{ab} = f_a(x) D_b(z)$ and $\overline{V}_{ff'}$ is defined in Eq.~\eqref{eq:Vffbar}.

\textit{Checks}:  We have erformed several checks to validate our results: At NLO, our findings are in complete agreement with the results of \cite{deFlorian:1997zj,deFlorian:2012wk}. 
In the soft plus virtual (SV) limit, where double-distribution terms alone contribute, the contributions of the $qq,\overline{qq}$ channel for NC and $\um\bm,\overline{\bm \um}$ channel for CC survive and they are found to be identical.  
Recently, NNLO QCD predictions for both NC and CC processes have been published~\cite{Bonino:2025tnf,Bonino:2025qta}. 
We find complete analytical agreement for all terms containing distributions for all the channels and numerical agreement for the regular parts.  We have expressed all the CFs in terms of single-valued polylogarithms. 
This allows to perform numerical computation over the entire kinematic range of $x'$ and $z'$.  

{\it Phenomenlogy:} We now illustrate the numerical impact of the NC and CC contributions to the unpolarized SF $\mathcal{F}_1$ as functions of the scaling variables $x$  in Figs.~\ref{fig:F1NC_x} and \ref{fig:F1CC_x}, at a center-of-mass energy $\sqrt{s} = 140~\mathrm{GeV}$ anticipated at the EIC.
NNPDF31 PDFs~\cite{NNPDF:2017mvq} and NNFF10 FFs~\cite{Bertone:2017tyb} are used consistently at LO, 
NLO, and NNLO, with $n_f = 4$ active quark flavors.

In Figs .~\ref{fig:F1NC_x}  and ~\ref{fig:F1CC_x} , $y$ is integrated over $[0.5,0.9]$ and $z$ over $[0.2,0.85]$. 
The average momentum transfer is defined as 
$Q_{\rm avg}^2 = s\,x\,y_{\rm avg}$, with the central scale choice 
$\mu_R = \mu_F = Q_{\rm avg}$. 
$\mathcal{F}_1^{\mathrm{NC}}$ and $\mathcal{F}_1^{\mathrm{CC}}$. 
In order to understand the dependence on the choice of scales for various values of $Q^2$, or equivalently values of $y = Q^2/(x s)$, 
we present in Figs.~\ref{fig:F1NC_x} and \ref{fig:F1CC_x} the variation of the renormalization and factorization scales 
$\mu_{R}$ and $\mu_{F}$ for $\mathcal{F}_1^{\rm{NC}}$ and $\mathcal{F}_1^{\rm{CC}}$ as functions of $x$. 
We choose six different values of $Q^2$, namely $Q^2_{avg}=\{30, 60, 100, 200, 400, 800\}$~GeV.  
The allowed range of $x$ for a given $Q^2$ is obtained by demanding $y$ to be in the range between 0.5 and 0.9.  
In Figs.~\ref{fig:F1NC_x} and \ref{fig:F1CC_x}, scales are varied independently by a factor two, i.e.\, $\mu_{R}^2$, $\mu_{F}^2$ $\in [Q^2/2, 2 Q^2]$ with the constraint $1/2 \leq \mu_R^2/\mu_F^2 \leq 2$ in the so-called seven-point scale variation. 

The large uncertainty bands of the LO predictions, in particular for the $\mathcal{F}_1^{\mathrm{NC}}$ and $\mathcal{F}_1^{\mathrm{CC}}$ SF, are due to the significant factorization scale dependence of the PDFs and FFs.
The plots in Figs.~\ref{fig:F1NC_x} and \ref{fig:F1CC_x} clearly demonstrate how higher order contributions decrease the $\mu_F$ dependence.  
The $\mu_R$ dependence can be assessed in a meaningful way starting from NLO, cf.\ plots on the left and in the center, and a comparison of uncertainty bands for the NLO and NNLO predictions show a significant reduction of the $\mu_R$ dependence in the latter case. 
Together with the observed apparent convergence of perturbative series for SIDIS, the reduction of the scale sensitivity give support to the robustness of our theoretical framework.

We now illustrate the numerical impact of the NC and CC contributions to the unpolarized SF $\mathcal{F}_1$ as functions of the scaling variable $z$  in Figs.~\ref{fig:F1NC_z} and \ref{fig:F1CC_z}, at a center-of-mass energy $\sqrt{s} = 140~\mathrm{GeV}$ anticipated at the EIC.
NNPDF31 PDFs~\cite{NNPDF:2017mvq} and NNFF10 FFs~\cite{Bertone:2017tyb} are used consistently at LO, 
NLO, and NNLO, with $n_f = 4$ active quark flavors.
For these plots, $x$ is integrated over $[0.1,0.8]$ and $y$ over $[0.5,0.9]$. 
The average momentum transfer is defined as 
$Q_{\rm avg}^2 = s\,x_{\rm avg}\,y_{\rm avg}$, with the central scale choice 
$\mu_R = \mu_F = Q_{\rm avg}$.

The shaded bands in all panels correspond to the seven-point variation of the 
renormalization and factorization scales, 
$\mu_R^2, \mu_F^2 \in [Q_{\rm avg}^2/2, 2\,Q_{\rm avg}^2]$ with the constraint 
$1/2 \le \mu_R^2 / \mu_F^2 \le 2$, capturing the combined uncertainty from 
scale choices after integration over $x$ and $y$ in different ranges, such that each range gives a fixed $Q_{avg}$, as indicated in each plot. The bands in the plots denote the seven-point scale variation around $Q^2_{avg}$ and, again, the sensitivity to the renormalization and factorization scales decreases significantly as we include the NLO and NNLO QCD corrections.
The results show that NNLO QCD corrections are significant and play a crucial role 
in reducing theoretical uncertainties from the choice of $\mu_R$ and $\mu_F$. 
The scale variation bands shrink notably from LO to NLO and NNLO, 
demonstrating improved perturbative convergence and enhanced stability at higher orders. 
At NNLO accuracy, the perturbative expansion exhibits both apparent convergence 
and robustness against scale variations, providing a reliable baseline for 
precision predictions of SIDIS observables.
Figs.~\ref{fig:F1NC_x} and \ref{fig:F1NC_z} show that the photon-mediated contribution dominates in the NC channel. 
This behavior can be understood from the propagator structure: the photon contribution scales as $1/Q^2$, whereas the $Z$-boson contribution is suppressed by the massive propagator $1/(Q^2 + M_Z^2)$, which reduces its impact for $Q^2 \ll M_Z^2$. In addition, the $Z$-boson couplings to fermions involve both vector and axial-vector components with weaker effective strengths compared to the electromagnetic coupling, further suppressing its contribution.

Similarly, Figs.~\ref{fig:F1CC_x} and \ref{fig:F1CC_z} indicate that the $W$-boson mediated contribution in the CC channel is suppressed. This suppression arises from the massive propagator $1/(Q^2 + M_W^2)$, as well as the chiral (purely left-handed) structure of the $W$-boson couplings, which restricts the contributing fermionic configurations and reduces the overall magnitude relative to massless photon exchange.

\emph{Summary}: 
We present the computation of NNLO QCD corrections to unpolarized and polarized  SIDIS, including both neutral-current and charged-current electroweak interactions.
We perform a phenomenological analysis for kinematics relevant to the Electron-Ion Collider (EIC) at its center-of-mass energy. By constructing polarization and CC/NC asymmetries, electroweak effects can be isolated and quantified. The different channels exhibit distinct sensitivities to the underlying partonic structure, providing complementary constraints on flavor-dependent fragmentation functions.
Our results establish a framework for NNLO-accurate studies of parton distributions and fragmentation functions in SIDIS. Combined with the broad kinematic coverage and particle identification capabilities of the EIC, they open the door to precision tests of QCD and electroweak dynamics in semi-inclusive processes.
These results can be utilized to enhance the extraction of PDFs and FFs, providing more precise, flavor-sensitive constraints. They provide an important check on recent independent computations ~\cite{Bonino:2025bqa,Bonino:2025qta, Bonino:2025tnf} and will enable high-precision theoretical predictions. Moreover, they will support ongoing and future studies of PDFs and the proton spin structure at the upcoming EIC.
%

\begin{widetext}

\begin{figure}[t]
\includegraphics[width=0.97\textwidth]{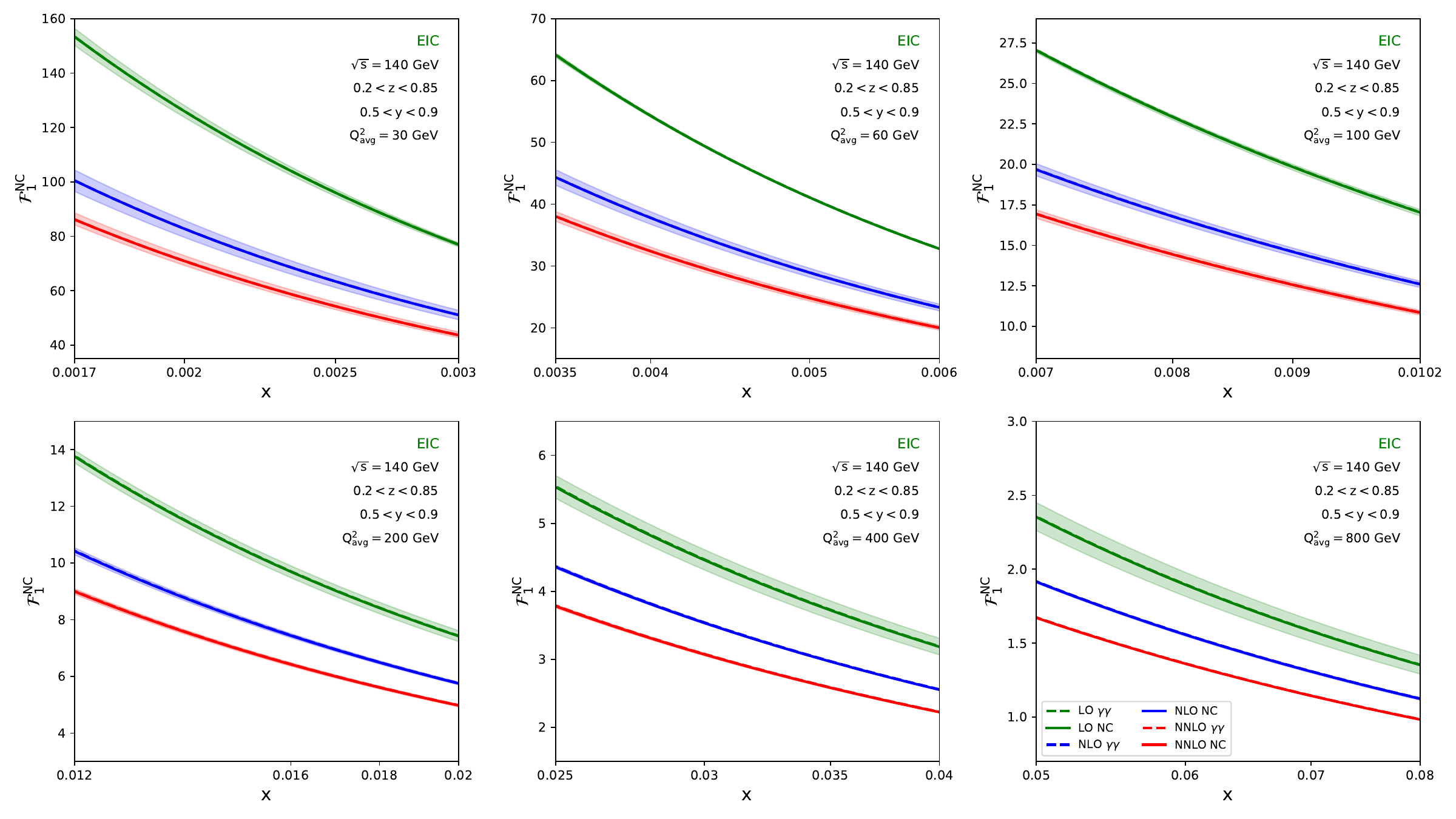}
\caption{Scale variation of $\mathcal{F}_1^{\mathrm{NC}}$ as a function $x$ for 6 different values of $Q^2$.  
}
\label{fig:F1NC_x}
\end{figure}

\begin{figure}[t]
\includegraphics[width=0.97\textwidth]{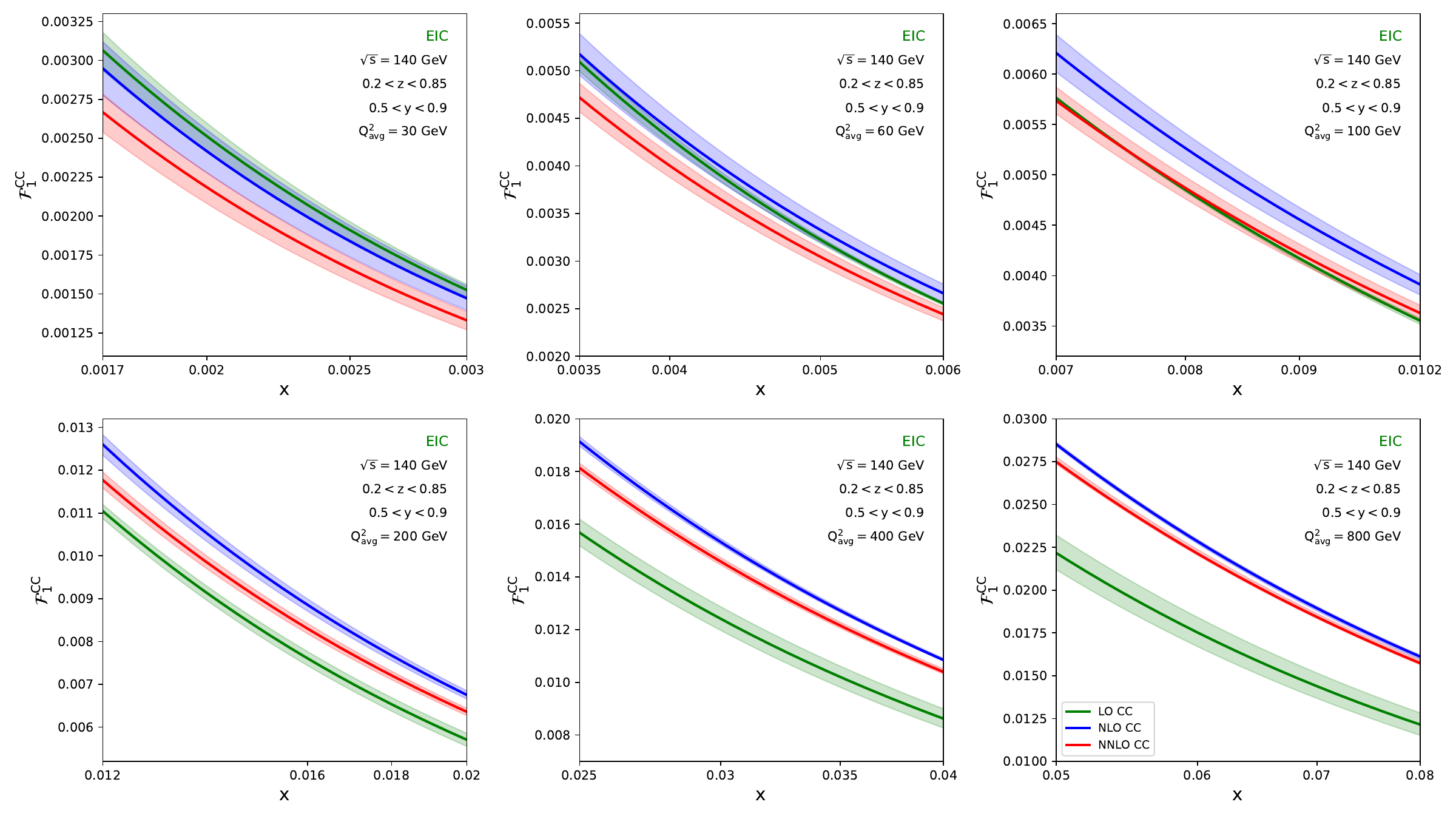}
\caption{Scale variation of $\mathcal{F}_1^{\mathrm{CC}}$ as a function $x$ for 6 different values of $Q^2$.  
}
\label{fig:F1CC_x}
\end{figure}

\begin{figure}[t]
\includegraphics[width=0.97\textwidth]{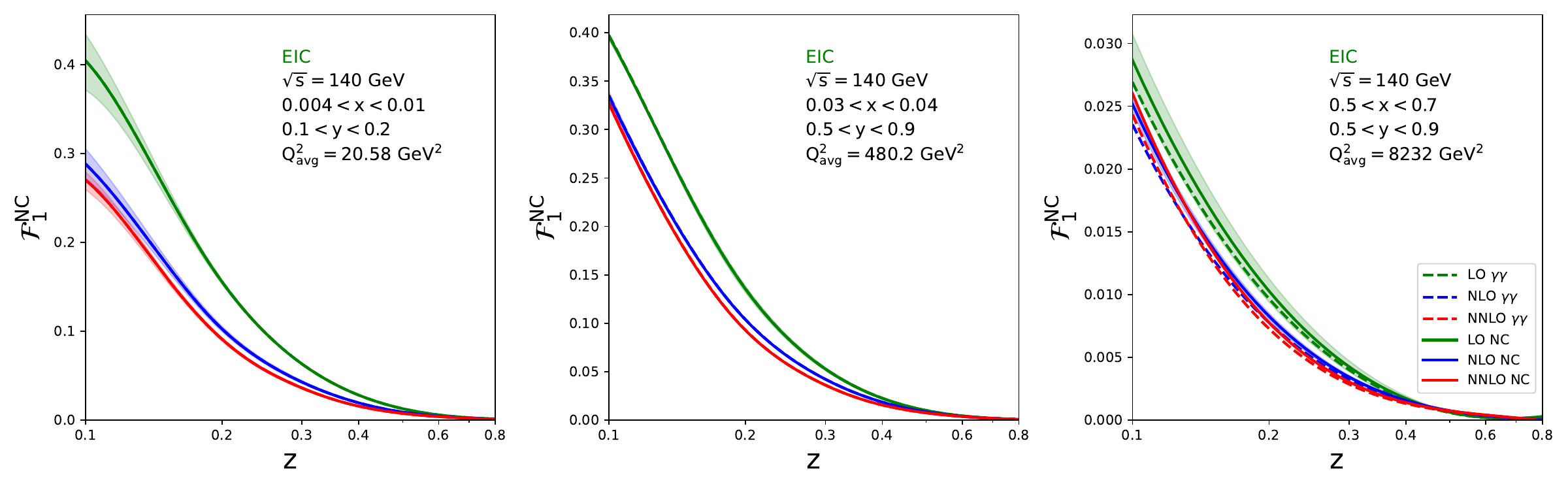}
\caption{Scale variation of $\mathcal{F}_1^{\mathrm{NC}}$ as a function $z$ for 3 different values of $Q^2$.  
}
\label{fig:F1NC_z}
\end{figure}

\begin{figure}[t]
\includegraphics[width=0.97\textwidth]{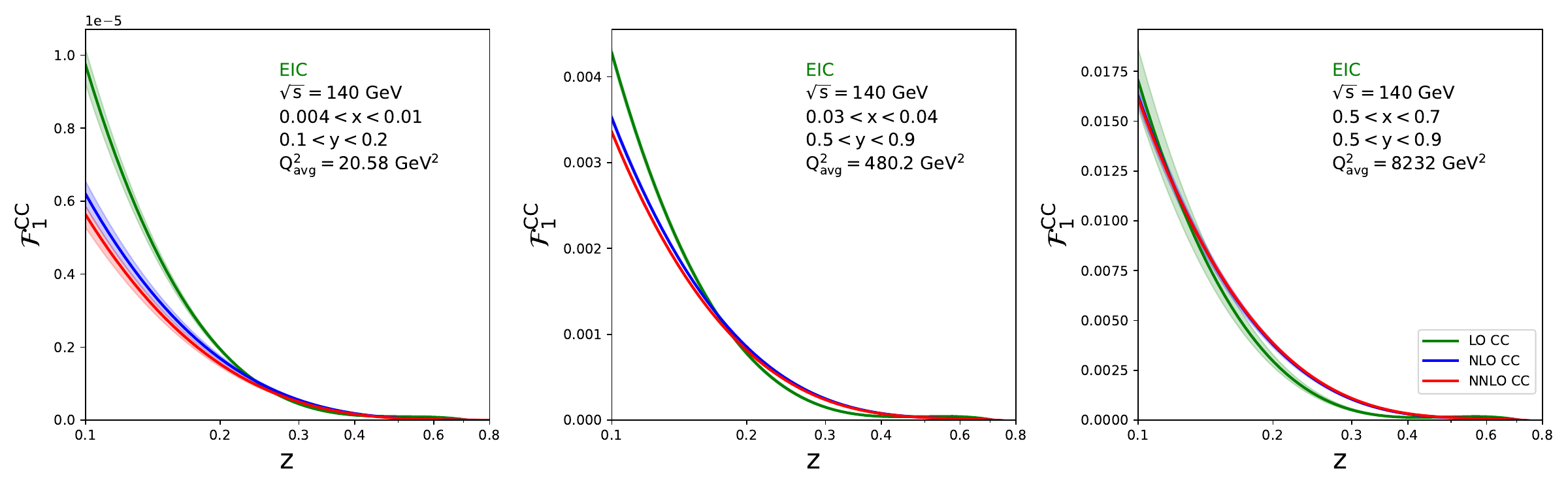}
\caption{Scale variation of $\mathcal{F}_1^{\mathrm{CC}}$ as a function $z$ for 3 different values of $Q^2$.  
}
\label{fig:F1CC_z}
\end{figure}

\end{widetext}
\begin{acknowledgments}
\emph{Acknowledgements:}  
We thank Roman N. Lee and Narayan Rana for usefull discussions.
This work has been supported through a joint Indo-German research grant by
the Department of Science and Technology (DST/INT/DFG/P-03/2021/dtd.12.11.21). 
S.M. acknowledges the ERC Advanced Grant 101095857 {\it Conformal-EIC}. 
In addition we would like to thank the computer administrative unit of IMSc for their help and support.
\end{acknowledgments}
\bibliographystyle{apsrev}
\bibliography{main}

\end{document}